# Magnetic properties and charge transport mechanisms in oxygen-deficient $Hf_xZr_{1-x}O_{2-y}$ nanoparticles

Oleksandr S. Pylypchuk[1], Eugene A. Eliseev[2], Andrii V. Bodnaruk[1], Valentin V. Laguta[2,3], Yuri O. Zagorodniy[2], Andrei D. Yaremkevych[1], Denis O. Stetsenko[1], Oksana V. Leshchenko[2], Victor N. Pavlikov[2], Lesya Demchenko[4,5], Victor I. Styopkin[1], Myroslav. V. Karpets[2,5], Olena M. Fesenko[1*], Victor V. Vainberg[1†], and Anna N. Morozovska[1‡]

[1] Institute of Physics of the National Academy of Sciences of Ukraine,46, Nauky Avenue, 03028 Kyiv, Ukraine

[2] Frantsevich Institute for Problems in Materials Science of the National Academy of Sciences of Ukraine, 3, str. Omeliana Pritsaka, 03142 Kyiv, Ukraine

[3]Institute of Physics of the Czech Academy of Sciences, Na Slovance 1999/2, 18200 Prague 8, Czech Republic

[4] Stockholm University, Department of Chemistry, Sweden

[5] Ye. O. Paton Institute of Materials Science and Welding, National Technical University of Ukraine "Igor Sikorsky Kyiv Polytechnic Institute", 37, Beresteisky Avenue, Kyiv, Ukraine, 03056

## Abstract

Study of nanoscale hafnia-zirconia physical properties is the key topic in fundamental and applied science. However, charge transport mechanisms and magnetic properties of $Hf_xZr_{1-x}O_2$ nanoparticles are very poorly studied both theoretically and experimentally. In this work we observed a superparamagnetic-like and superparaelectric-like response of ultra-small $Hf_xZr_{1-x}O_{2-y}$ nanoparticles prepared by the solid-state organonitrate synthesis. The EPR spectra of $Hf_xZr_{1-x}O_{2-y}$ nanopowders reveal the presence of paramagnetic defect centers, which may be hafnium and/or zirconium ions, which trapped an electron near an oxygen vacancy and changed their valence state from the non-paramagnetic +4 to the paramagnetic +3 state. The Raman spectra indicate the decisive role of surface defects, presumably oxygen vacancies, for all studied x = 1, 0.6, 0.5 and 0.4 At the same time the EELS analysis does not reveal any noticeable concentration of magnetic impurities in the $Hf_xZr_{1-x}O_{2-}$

[*] Corresponding author: fesenko.olena@gmail.com
[†] Corresponding author: viktor.vainberg@gmail.com
[‡]Corresponding author: anna.n.morozovska@gmail.com

$_y$ nanopowders, and the X-ray diffraction analysis reveals the dominant presence of the orthorhombic phase. We observed that the quasi-static relative dielectric permittivity of the $Hf_xZr_{1-x}O_{2-y}$ nanopowders overcomes $10^6$ - $10^7$ and related the colossal values with the superparaelectric state of the nanoparticles' cores induced by the flexo-electro-chemical strains. It has been found that ultra-small $Hf_xZr_{1-x}O_{2-y}$ nanoparticles reveal posistor effect and relatively large values of accumulated charge. Thus, obtained results open the way for creation of silicon-compatible ferroics – oxygen-deficient $Hf_xZr_{1-x}O_{2-y}$ nanoparticles with superparamagnetic and superparaelectric properties, which may be used in advanced FETs and electronic logic elements.

## I. INTRODUCTION

Thin $Hf_xZr_{1-x}O_2$ $(0.4 \leq x \leq 0.6)$ films, their multilayers and heterostructures are indispensable silicon-compatible ferroelectric materials for advanced electronic memories [1, 2] and logic devices [3, 4]. Experimental and theoretical studies of nanoscale hafnia-zirconia oxides physical properties are pivotal in fundamental and applied science [1-4]. However, polar properties of $Hf_xZr_{1-x}O_2$ nanoparticles are poorly studied theoretically, as well as direct experimental observations of their ferroelectricity are still absent. To the best of our knowledge, the charge transport mechanisms and magnetic properties of $HfO_2$ nanoparticles are unexplored. Also, the influence of zirconium doping on the magnetic properties of nanoscale $HfO_2$ has not been studied. At the same time, addition of zirconium has a very strong stabilizing influence on the ferroelectricity revealed in thin $Hf_xZr_{1-x}O_2$ films [5, 6], $HfO_2$-$ZrO_2$ multilayers [7] and superlattices [8].

A room-temperature ferromagnetism in thin $HfO_2$ films, which are nonmagnetic in bulk, was revealed by Venkatesan et al. [9], Coey et al. [10] and Hong et. al. [11, 12]. At that the concentration of magnetic impurities in the magnetic thin $HfO_2$ films [9-12], as well as in other oxides, such as $SnO_2$ nanosheets [13] and thin films [14], was much smaller than $10^{-2}$ wt.%. Since annealing in oxygen ambient destroys or greatly reduces the ferromagnetic response, it was concluded that the oxygen vacancies concentrated at the film surfaces or/and interfaces are the main source of the room-temperature magnetization and emerging *surface-induced* magnetism [15, 16, 17].

The physical mechanisms responsible for emergence of ferromagnetic, ferroelectric and/or antiferroelectric properties in nanoscale $Hf_xZr_{1-x}O_2$ with sizes from 5 to 30 nm are still unclear [18, 19]. It is probably related with the dominant role of surface and size effects [20, 21], field-induced effects [22] and/or competing phases, which determine the spontaneous polarization switching [23, 24]. The origin of the size-induced ferroelectricity in the nanoscale $Hf_xZr_{1-x}O_2$ is related to the transition from the non-polar monoclinic m-phase (space group $P2_1/c$) to the polar orthorhombic o-phase (space group $Pca2_1$), which is metastable compared to the bulk m-phase. Following recent experimental observations [25], the transition path may be indirect, going through the non-polar tetragonal t-phase. However, it is hardly possible to separate the t-phase (space group *$P4_2/nmc$*) and o-phases (space groups *Pbca*, *Pbcm* and ferroelectric *Pca21*) in small $Hf_xZr_{1-x}O_2$ nanoparticles using X-ray diffraction analysis (XRD), because corresponding peaks are diffused and/or very close to each other [26].

Several experimental [27, 28] and theoretical [29, 30] works revealed a leading role of the oxygen vacancies [31, 32] in the appearance and stabilization of the o-phase in nanoscale $Hf_xZr_{1-x}O_2$. In particular, it was predicted theoretically [29] that the ferroelectric phase can be induced in thin $HfO_2$ films by the "flexo-electro-chemical" coupling. The coupling is the joint action of the omnipresent flexoelectric effect [33, 34] that is especially important at nanoscale [35, 36], electrostriction and chemical pressure [37]. Due to the flexo-electro-chemical coupling, the reversible ferroelectric polarization, as high as (5 – 30) $\mu C/cm^2$, can be induced by the oxygen vacancies in $HfO_2$ films of thickness less than 10 – 30 nm [29]. Later on, the Landau-Ginzburg-Devonshire (LGD) thermodynamic approach [38], density functional theory (DFT) calculations [39, 40] and dielectric measurements [41] reveal that oxygen-deficient $Hf_xZr_{1-x}O_{2-y}$ nanoparticles could exhibit ferroelectric-like properties [38-40], such as a colossal dielectric response in a wide frequency range [41], as well as demonstrate resistive switching and pronounced charge accumulation [42].

Inspired by these results, in this work we study charge transport, magnetic and dielectric properties of small (~ 5 – 10 nm) oxygen-deficient $Hf_xZr_{1-x}O_{2-y}$ nanoparticles (x = 1 – 0.4) prepared by the solid-state organonitrate synthesis. We observed a superparamagnetic-like (SPM) response of the $Hf_xZr_{1-x}O_{2-y}$ nanoparticles. We also observed that the relative dielectric permittivity of the $Hf_xZr_{1-x}O_{2-y}$ nanopowders overcomes $10^6$ in the static regime and related the colossal values with the superparaelectric (SPE) states of the ultra-small cores of the nanoparticles. Therefore, obtained results open the way to create innovative silicon-compatible multiferroics – oxygen-deficient $Hf_xZr_{1-x}O_{2-y}$ nanoparticles with the SPM and SPE properties.

## II. SAMPLES PREPARATION AND CHARACTERIZATION

Oxygen-deficient $Hf_xZr_{1-x}O_{2-y}$ (x = 1, 0.4., 0.5 and 0.6) nanopowders were prepared by the solid-state organonitrate synthesis from the mixtures of zirconium and hafnium nitrate salts, $ZrO(NO_3)_2$ $2H_2O$ and $Hf(NO_3)_2$ $2H_2O$, dissolved in distilled water to the concentration of 5 – 10 % or less (see details in Refs. [38-41]). As-prepared samples were subjected to heat treatment in the CO + $CO_2$ ambient at 500 – 600°C for several hours. In result, the expected oxidation degree appeared high enough to modify optical properties of the nanopowders (the samples become black).

The XRD analysis of the nanopowders confirmed a coexistence of the monoclinic m-phase (space group P21/c), whose the content varies from 13 to 5 wt. %, and three inseparable orthorhombic o-phases (space groups Pbca, Pbcm and ferroelectric Pca21), whose content varies from 87 to 96 wt. %, inside the nanoparticles (see **Fig. S1** and **Table S1** in **Supplement S1**, Supplementary Materials).

The average size (~8 – 10 nm) of the $Hf_xZr_{1-x}O_{2-y}$ nanoparticles was determined by the transmission electron microscopy (TEM). The dispersion of the particle sizes from 3 to 30 nm was determined approximately from multiple TEM images. The average size depends slightly on the Hf/Zr ratio and correlates with the size of coherent scattering regions (CSR) determined by the XRD (see **Table S1** in Supplementary Materials). The nanoparticles have a rounded quasi-spherical shape (see the top and two middle rows in **Fig. 1(a)-(d)** as typical examples). Important, that the nanoparticles have a pronounced crystalline structure (see the top row in **Fig. 1**) and the dispersion of their sizes is relatively small (see the third row in **Fig. 1**). However, it is seen that the $Hf_{0.5}Zr_{0,5}O_{2-y}$ nanoparticles attract each other (probably due to electric charging), in contrast to the $Hf_xZr_{1-x}O_{2-y}$ nanoparticles with x = 1, 0.4 and 0.6.

To obtain more information about the samples morphology and atomic contrast, the $Hf_xZr_{1-x}O_{2-y}$ nanoparticles were studied by the scanning electron microscopy (SEM). The SEM in backscattered electron mode gives information about the average atomic number $Z_{aver}$ of material (Z-contrast). The magnitude of $Z_{HfO2}$ is almost two times larger compared to that of $Z_{ZrO2}$, and therefore these materials can be easily discerned in the Z-contrast mode. The analysis shows that $Hf_xZr_{1-x}O_{2-y}$ samples are mostly homogeneous (see bottom images in **Fig. 1**). The difference in signals in some areas is related to the difference in thickness. The values of $Z_{aver}$ obtained from Z-contrast images are close to those calculated using the chemical composition formula of the material for all studied "x". The minimal size of the agglomerated particles was 100 – 300 nm for the most cases.

However, the $Hf_{0.5}Zr_{0.5}O_{2-y}$ sample charges strongly under irradiation by an electron probe, which precludes obtaining stable SEM images at magnifications of 30 – 40 thousand. It appeared possible to resolve particles with the minimal size of 300 – 600 nm. Due to the strong charging, it was impossible to study the morphology of the $Hf_{0.5}Zr_{0.5}O_{2-y}$ particles in more detail at significant magnifications. This result also indicates that $Hf_{0.5}Zr_{0.5}O_{2-y}$ samples have significantly lower conductivity compared with other samples. It is known that oxygen-deficient $ZrO_{2-y}$ exhibits significantly enhanced electrical conductivity**,** primarily due to the introduction of oxygen vacancies acting as charge carriers [43]. Thus, one may assume that $Hf_{0.5}Zr_{0.5}O_{2-y}$ samples contain less oxygen vacancies compared with other compositions, or the vacancies form some complexes with zero net charge (e.g., dipoles or multipoles) in the samples.

Larger plate-like agglomerated particles with a flattened shape were observed only in the $Hf_{0.4}Zr_{0.6}O_{2-y}$ samples; they are absent in other samples. The agglomerates are few microns in size and have a minimal thickness of 100 – 200 nm. Usually such plate-like particles are formed in materials with closely packed blocks, which have minimal surface energy. It should be noted that the minimal size of the $Hf_{0.4}Zr_{0.6}O_{2-y}$ particles resolved by SEM is 50 – 200 nm, which is slightly less than for other “x” and may be due to the presence of close-packed blocks.

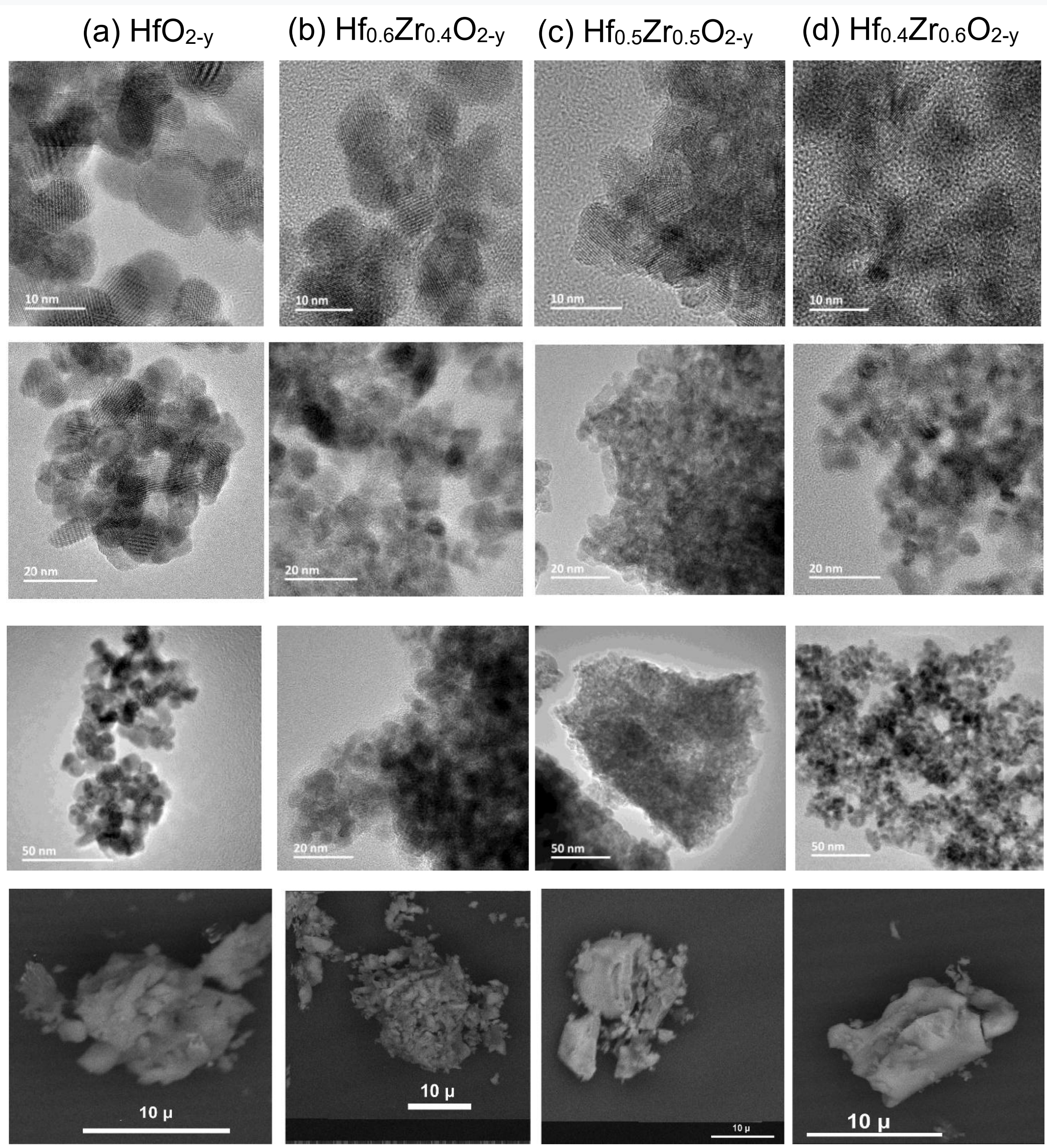

**FIGURE 1. (a-d)** Typical TEM (the top and two middle rows) and SEM (the bottom row) images of the $Hf_xZr_{1-x}O_{2-y}$ nanopowders.

Note that thoroughly performed chemical analysis by the energy-dispersive X-ray spectroscopy (EDS), and Electron Energy-Loss Spectroscopy (EELS) studies did not reveal any noticeable concentration of magnetic impurities (such as Mn, Fe or Co atoms) in the sintered $Hf_xZr_{1-x}O_{2-y}$ nanopowders (see **Supplement S2** in the Supplementary Materials and Refs. [39 - 42] for details). The estimated uncertainty of the Hf/Zr atomic ratio determined by TEM-EDS is approximately ±0.5 at.%, while the corresponding normalized metal-site fraction uncertainty is about ±0.005 (see **Fig. S2** and **Table S2** in **Supplement S2.A**). The EELS, being a volumetric method, can provide the information which value is analogous to the XPS analysis, while the XPS response is rather surface-related. The EELS spectrum, shown in **Fig. S3** in **Supplement S2.B**, is a typical oxygen O K-edge EELS spectrum of $Hf_{1-x}Zr_xO_2$ nanoparticles recorded in the TEM mode (on example of the $Hf_{0.35}Zr_{0.65}O_{2-y}$ sample). The O K-edge arises from transitions of O 1s core electrons into unoccupied O 2p states hybridized with the Hf 5d and Zr 4d orbitals and therefore probes the unoccupied metal *d*-derived electronic states. The spectrum exhibits four dominant features labeled as A, B, C and D. The EELS spectrum demonstrates the pre-edge appearing at the low-energy side of the peak A (at approximately 529 – 531 eV). The feature can be attributed to oxygen vacancy-induced defect states [44, 45, 46]. Oxygen vacancies introduce excess electrons, which partially occupy metal *d* states reducing the number of unoccupied $t_{2g}$ states, and, consequently, the intensity of the $t_{2g}$-related peak A.

Although the absolute values of the oxygen vacancies concentration cannot be directly quantified from the O K-edge EELS, relative changes in the concentration can be inferred from the intensity of defect-related pre-edge features and the redistribution of spectral weight between the A and B peaks [47, 48, 49, 50]. The O K-edge EELS provides a sensitive probe of oxygen vacancy-related electronic structure, but it does not allow direct determination of vacancy concentration without external calibration (either by the first-principles calculations of DFT-based simulated spectra, or by complementary experimental techniques). Any additional edges, corresponding to transition-metal impurities, such as Mn (640 – 651 eV), Fe (708 – 721 eV), or Co (779 – 794 eV) edges, are absent in the measured energy range, confirming the absence of these elements within the detection limits of EELS.

At the same time, the EPR spectra of $Hf_xZr_{1-x}O_{2-y}$ nanopowders confirm the presence of $nd^1$-paramagnetic defect centers. Typical EPR spectra of the $Zr_{0.5}Hf_{0.5}O_{2-y}$ nanopowders recorded at different temperatures are shown in **Fig. 2**. The spectra are dominated by an intense and broad resonance line centered at approximately 2800 G. With decreasing temperature, this resonance slightly shifts toward lower magnetic fields and becomes broader. The integral intensity of this spectral line remains almost unchanged upon cooling down to 40 K. These features indicate ferromagnetic resonance arising from spins ordered due to exchange interactions. We may assume that these spins can be attributed to paramagnetic $Hf(Zr)^{+3}$ – $V_O$ centers ($F^+$-type centers). Such centers can be formed when an electron is trapped by positively charged oxygen vacancy and is localized predominantly at the $5d^1/4d^1$ orbital of one of the four neighboring Hf(Zr) ions. The formation of such $F^+$-type centers in oxygen-deficient oxides has been studied in detail in Ref. [51], where very similar dielectric, magnetic and conduction properties have been observed. The microscopic model of such centers was presented in Ref. [52] on example of $BaTiO_3$.

The assumption is corroborated by the EPR spectrum, typical for isolated $Hf(Zr)^{+3}$ – $V_O$ centers, which are not exchange-coupled. Such spectrum was observed in the $Zr_{0.5}Hf_{0.5}O_2$ powder, which was not annealed in the CO + $CO_2$ ambient, in which the concentration of oxygen vacancies should be low (see bottom spectrum in **Fig. 2**). This EPR spectrum is characterized by narrow spectral lines originating from isolated spins (paramagnetic ions). In particular, the spectral line at ≈ 1600 G with g-factor ≈ 4.26 is attributed to the $Fe^{3+}$ ions in a strongly distorted local environment. The intense narrow line with g = 2.00 most likely originates from the radical species. The spectral feature in the range of magnetic field 3350 – 3400 (namely, two spectral lines with g-factors $g_1$ = 2.004 and $g_2$ = 1.956 – 1.977 shown in the inset to **Fig. 2**) are assigned to $Zr^{3+}$ and/or $Hf^{3+}$ formed near oxygen vacancies, as previously reported in Ref.[42].

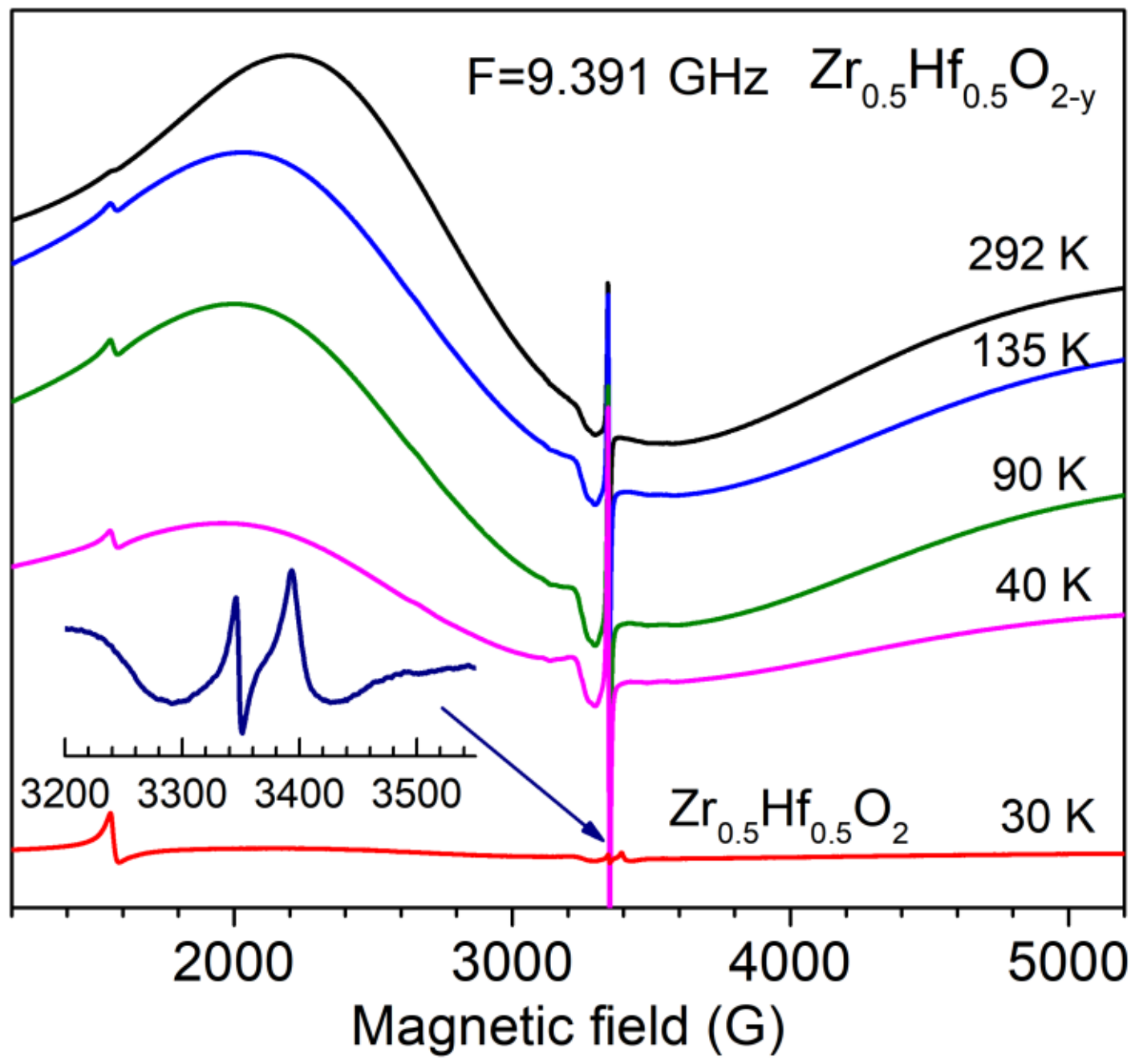

**FIGURE 2.** EPR spectra of the $Zr_{0.5}Hf_{0.5}O_{2-y}$ nanopowder annealed in CO + $CO_2$ ambient. The spectra are measured at 292, 135, 90, and 40 K (black, blue, green and magenta curves, respectively). The EPR spectrum of the non-annealed $Zr_{0.5}Hf_{0.5}O_2$ powder measured at 30 K is shown by the red curve. The inset shows spectral lines, which may be ascribed to the $Hf(Zr)^{+3}$ – $V_O$ isolated centers.

In summary, the dominant EPR signal, observed in $Hf_{1-x}Zr_xO_{2-y}$ nanoparticles, originates primarily from clusters enriched by paramagnetic species. The concentration of these defects is sufficient to induce a spin-ordered state that is confirmed by ferromagnetic resonance (FMR) spectral feature in the EPR spectrum. We assumed that such species can be $Hf(Zr)^{+3}$ – $V_O$ centers. We also note that the $Hf(Zr)^{+3} - V_O$ center exhibits an electric dipole moment, which should contribute to the polar properties of the $Hf_{1-x}Zr_xO_{2-y}$ nanoparticles. The polar properties are considered in Section III.

The Raman spectra of $Hf_xZr_{1-x}O_{2-y}$ nanopowders, shown in **Fig. 3**, have a broad band in the 1000 – 3000 cm$^{-1}$ range, which is typical for the samples containing a high concentration of Raman-active luminescence centers associated with defects, presumably oxygen vacancies [39, 40] (see **Table S3** in **Supplement S3**). The broad band intensity decreases with increase in zirconium content, which correlates with the stabilization of the more homogeneous o-phase and may indicate a decrease in the number of oxygen vacancies. For the samples with x = 0.4, the intensity of this band increases slightly in comparison with its intensity for x = 0.5, which we relate to the increase in the m-phase fraction and the associated growth in the number of luminescence centers. No distinct phonon modes were observed below 350 cm$^{-1}$, which is consistent with the strongly disordered nanosized structure of the powders. This result agrees with the XRD data, which reveals a coexistence of the o-phase and m-phase in the studied nanoparticles.

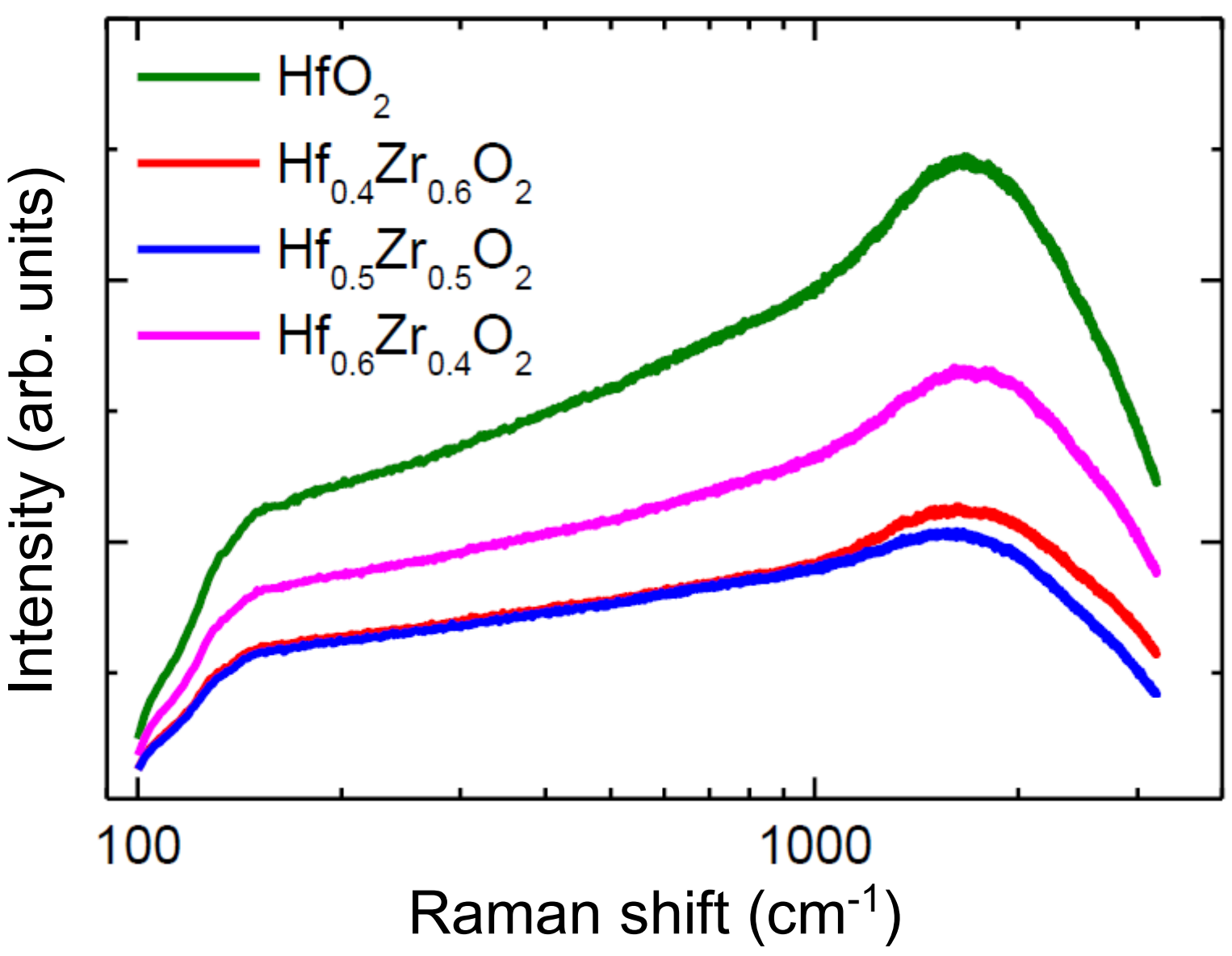

**FIGURE 3. (a)** Raman spectra of the $Hf_xZr_{1-x}O_{2-y}$ nanopowders with different hafnium content x = 1, 0.6, 0.5 and 0.4.

## III. MAGNETIC, DIELECTRIC AND CHARGE TRANSPORT PROPERTIES OF $Hf_xZr_{1-x}O_{2-y}$ NANOPOWDERS

### A. Measurements of the magnetic response and its analysis

Magnetostatic measurements of $Hf_xZr_{1-x}O_{2-y}$ nanopowders were performed using the LDJ 9500 magnetometer with a vibrating sample. The powders were poured into a polypropylene capsule and tightly packed to prevent the particles from “bouncing”, as the sample vibrates with an acceleration of 25 g (g is the acceleration of gravity), because the bouncing could distort experimental results.

Magnetization of $Hf_xZr_{1-x}O_{2-y}$ nanopowders was measured in the temperature range from 100 K to 350 K, and the SPM-like behavior was observed entire the range. Magnetization curves of the $Hf_xZr_{1-x}O_{2-y}$ nanopowders, measured at the room temperature, are shown in **Fig. 4(a)**. It is seen that the curves have a pronounced SPM behavior, and the value of saturated magnetization $M_s$ monotonically decreases from $3.06 \cdot 10^{-2}$ emu/g to $0.66 \cdot 10^{-2}$ emu/g with a decrease in hafnium content "x" from 1 to 0.4 (see different curves in **Fig. 4(a)**). This tendency can be explained by the gradual decrease in the magnetic defect concentration with decrease in the hafnium content "x". This tendency agrees with the decrease in the intensity of Raman spectra (compare green, magenta and blue curves in **Fig. 3**), which indicate the leading role of the oxygen vacancies in the formation of the SPM response.

While the increase of zirconium content decreases the saturated magnetization, the influence of zirconium doping on $M_s$ also saturates for $x < 0.5$ (compare blue and red curves in **Fig. 4(a)**, which are very close), indicating on the possible role of magnetic defects accumulated near the surface. This trend agrees with the intensity and FWHM of Raman spectra (compare blue and red curves in **Fig. 3(a)**, which are very close too). Schematic illustration of the magnetic percolation [53] emerging between the paramagnetic centers, which may be responsible for the appearance of SPM properties in the $Hf_xZr_{1-x}O_{2-y}$ nanoparticles is shown in **Fig. 4(b)**. Here the radii of spheres designate the exchange radius of magnetic defects, and percolation starts when the infinite cluster appears at the particle surface.

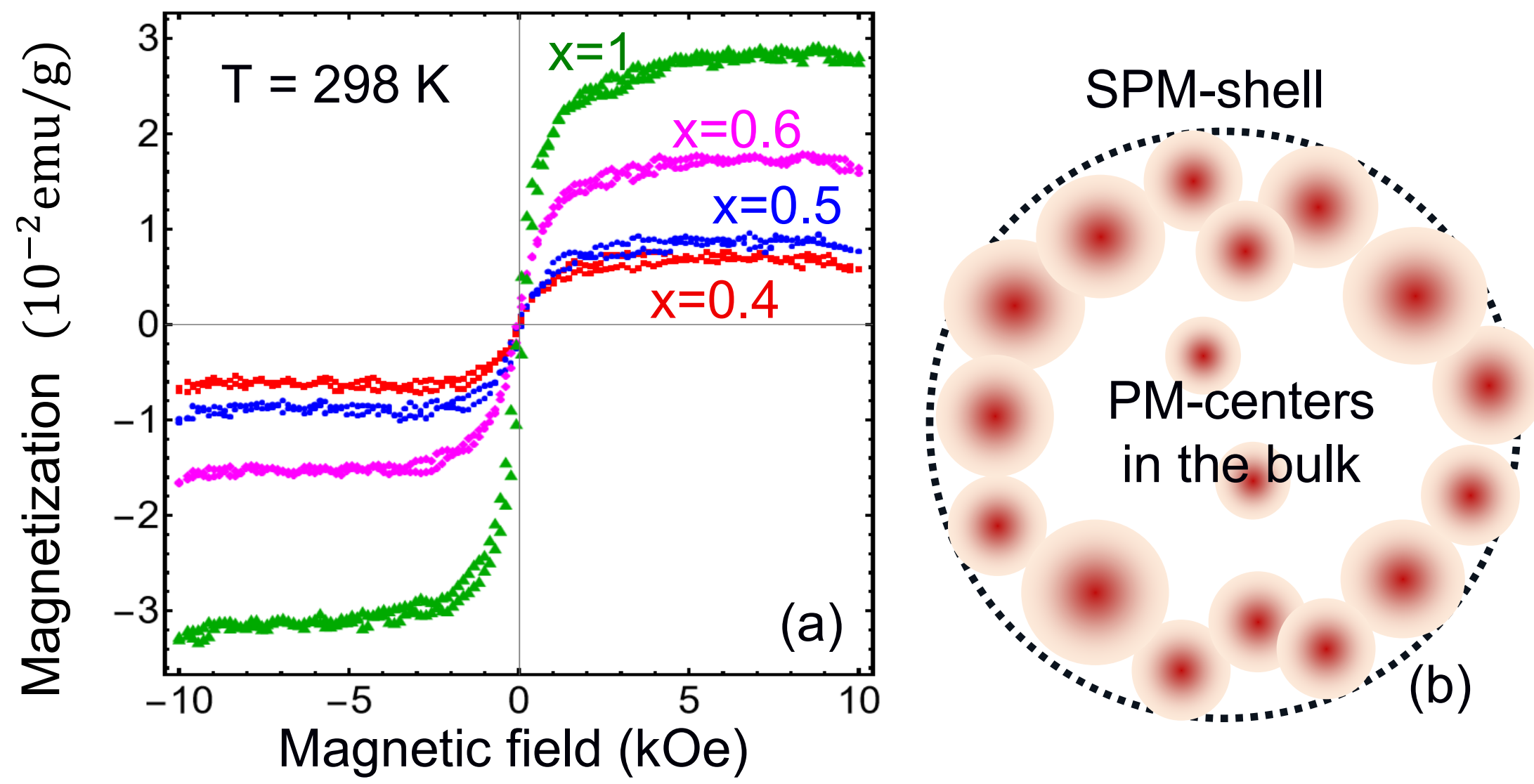

**FIGURE 4. (a)** Magnetization curves of the $Hf_xZr_{1-x}O_{2-y}$ nanopowders measured at room temperature $T$ = 298 K. **(b)** Schematic illustration of the magnetic percolation at the surface, and its absence in the bulk of the nanoparticle.

The dependences of the magnetization versus the magnetic field measured at 100, 160, 220, 280 and 340 K are shown in **Figs. 5(a)-(d)** for different hafnium content “x”. It is seen that the difference between the forward and backward directions of the magnetic field sweep is the largest at x = 0.4 and 0.5, much smaller at x = 0.6 and almost absent at x = 1. The backward magnetization curves are closer to superparamagnetic curves, while the forward curves acquire a negative slope with increase in magnetic field at x = 0.4 and x = 0.5. It is unclear why the difference between the forward and the backward sweeps is pronounced at x = 0.4 and 0.5 only. To quantify observed features of the magnetic response, below we simulate the magnetic properties of $Hf_xZr_{1-x}O_2$ nanopowders.

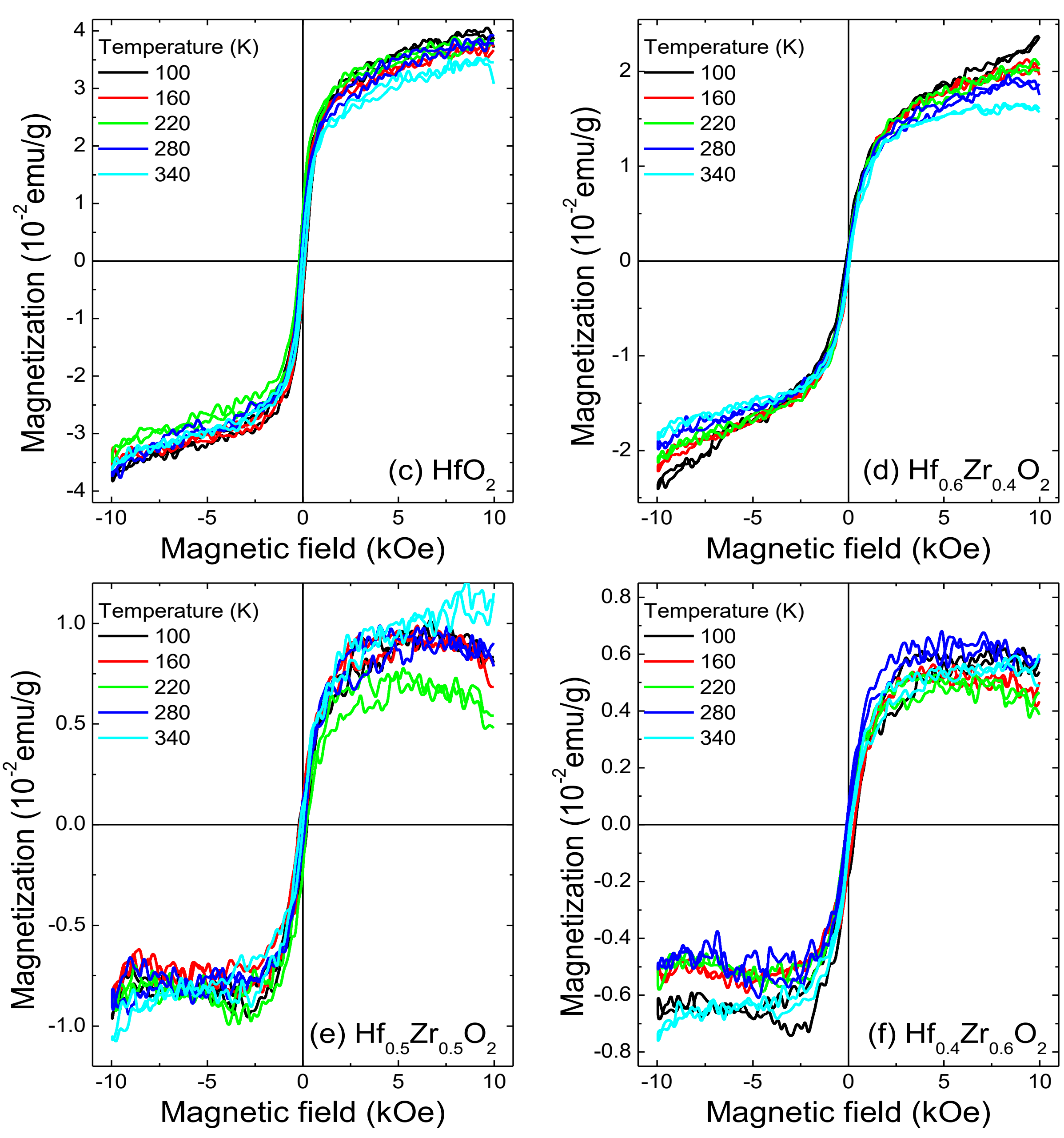

**FIGURE 5.** Dependence of the magnetization vs. the magnetic field applied to the $Hf_xZr_{1-x}O_{2-y}$ nanopowders with different hafnium content x = 1 **(a)**, 0.6 **(b)**, 0.5 **(c)** and 0.4 **(d)**. Different curves correspond to the temperatures 100, 160, 220, 280 and 340 K (as indicated in legends).

To simulate the magnetic response of $Hf_xZr_{1-x}O_2$ nanopowders, let us consider a mixture consisting of ultra-small superparamagnetic (SPM) nanoparticles and larger paramagnetic (PM) nanoparticles. Also, the SPM inclusions can appear inside the larger particles. It was shown by Korenblit and Shender [54] that, according to percolation theory, long-range magnetic order in the system occurs as soon as the infinite cluster forms. The distance between the magnetic centers, which form an infinite cluster, and therefore, fulfill the percolation condition ("percolation threshold"), is defined as a percolation radius. These speculations allow us to assume the following dependence of the magnetization $M$ on the applied quasi-static magnetic field $H$:

$$M(H) = M_{PM}(H) + M_{SPM}(H) \approx \chi_m H + \int_{\mu_{min}}^{\mu_{max}} F(\mu - \mu_S) M_P(\mu) f\left(\frac{\mu H}{k_B T}\right) d\mu + \delta M, \quad (1)$$

where the first term is the PM (or diamagnetic) contribution, and the second one can be the SPM contribution in the nanoparticle ensemble. The value $\delta M$ can be nonzero reflecting a very small device-related systematic error (e.g., shift).

**PM contribution.** The linear approximation for the PM (or diamagnetic) contribution,

$$M_{PM}(H) \approx \chi_m H, \quad (2)$$

is valid for magnetic fields $|H| << k_B T / g\mu_B S$, where $\mu_B$ is the Bohr magneton, $k_B$ is the Boltzmann constant, $g$ is the g-factor and $S$ is the total angular momentum quantum number (or "total spin"). From the estimation, the magnetic field should be much smaller than 800 kOe at room the temperature, that is true for our experiments. The paramagnetic permittivity is $\chi_m = \frac{C_{CW}}{T - \theta}$, where θ is the Curie temperature and $C_{CW}$ is the Curie-Weiss constant. The constant $C_{CW} = \frac{N_{PM}\mu_p^2}{3k_B} \approx \frac{4N_{PM}\mu_B^2 g^2 S(S+1)}{3k_B}$, where $N_{PM}$ is the number of paramagnetic spins $S$ in a unit mass volume.

**SPM contribution.** Following Binder and Young [55], and Wiekhorst et al. [56] the integration (or averaging) of the last term in Eq.(1) reflects the fact that the number of elementary spins, which contribute to the magnetic moment μ of a given SPM region, can be different in different nanoparticles. The moment μ can fluctuate around the average value $\mu_S(T)$ in dependence on the sharpness of its distribution function $F(\mu - \mu_S)$. The function is defined as $\mu_S(T) = \int_{\mu_{min}}^{\mu_{max}} F(\mu - \mu_S)\mu d\mu$. The magnetization amplitude $M_P(\mu) = N_S \mu$, where $N_S$ is the average number of spins $S$ in one gram of the material with the average magnetic moment $\mu_S(T)$. Depending on the $S$ value, one can use Langevin or Brillouin formulae for the function $f(x)$ describing the SPM contribution:

$$f(x)=\begin{cases}\coth(x)-\frac{1}{x}, & \text{Langevin function,}\\ \frac{2S+1}{2S}\coth\left(\frac{2S+1}{2S}x\right)-\frac{1}{2S}\coth\left(\frac{x}{2S}\right), & \text{Brillouin function.}\end{cases} \quad (3)$$

As one can see the Brillouin function transforms into the Langevin function in the limit $S >> 1$, since $\coth\left(\frac{2S+1}{2S}x\right)\approx\coth(x)$ and $\coth\left(\frac{x}{2S}\right)\approx\frac{2S}{x}$ with high accuracy for $S\geq 10^3$.

Assuming that all SPM particles are formed by the same magnetic defects, one can use the following expression for the fitting of experimental data:

$$M(H)\approx\chi_m H+M_S f(\tilde{\mu}\,H)\ +\delta M. \quad (4)$$

The reduced magnetic moment $\tilde{\mu}=\frac{\mu}{k_B T}$ has the dimensionality of 1/Oe, $M_S$ is the magnetization in saturation.

Fitted dependences of the room temperature magnetization vs. the magnetic field applied to the $Hf_xZr_{1-x}O_{2-y}$ nanopowders are shown in **Fig. 6(a)-(d)** by solid black curves. The fitting parameters are listed in **Table I.** Dependences of the fitting parameters vs. the hafnium content "x" are shown in **Supplement S4** (see Supplementary Materials). We obtained that $S>10^3$, which proves the SPM nature of the magnetic response. Under the fitting, we assumed that all nanoparticles are single-domain due to their small sizes (~5 – 10 nm), which are smaller than the width of the Bloch-type ferromagnetic wall.

It is seen from **Table I** that the negative slope of the forward magnetization curves is related to the opposite sign of the PM (or diamagnetic) and SPM contributions. Namely, the paramagnetic/diamagnetic constant $\chi_m<0$ for the forward sweeps, and $\chi_m>0$ for the backward sweeps at x = 0.4, 0.6 and 0.6 (see the third column in **Table I**). The difference between the forward and backward directions of the field sweep may be related to the presence of "stubborn" magnetic defects, which contribution is the most significant at x = 0.4 and 0.5, meanwhile the contribution of SPM dipoles (parameter $M_S$) is the smallest at x = 0.4 and 0.5. We may assume that the role of stubborn magnetic defects is the strongest in the case of maximal compositional disorder near x = 0.5.

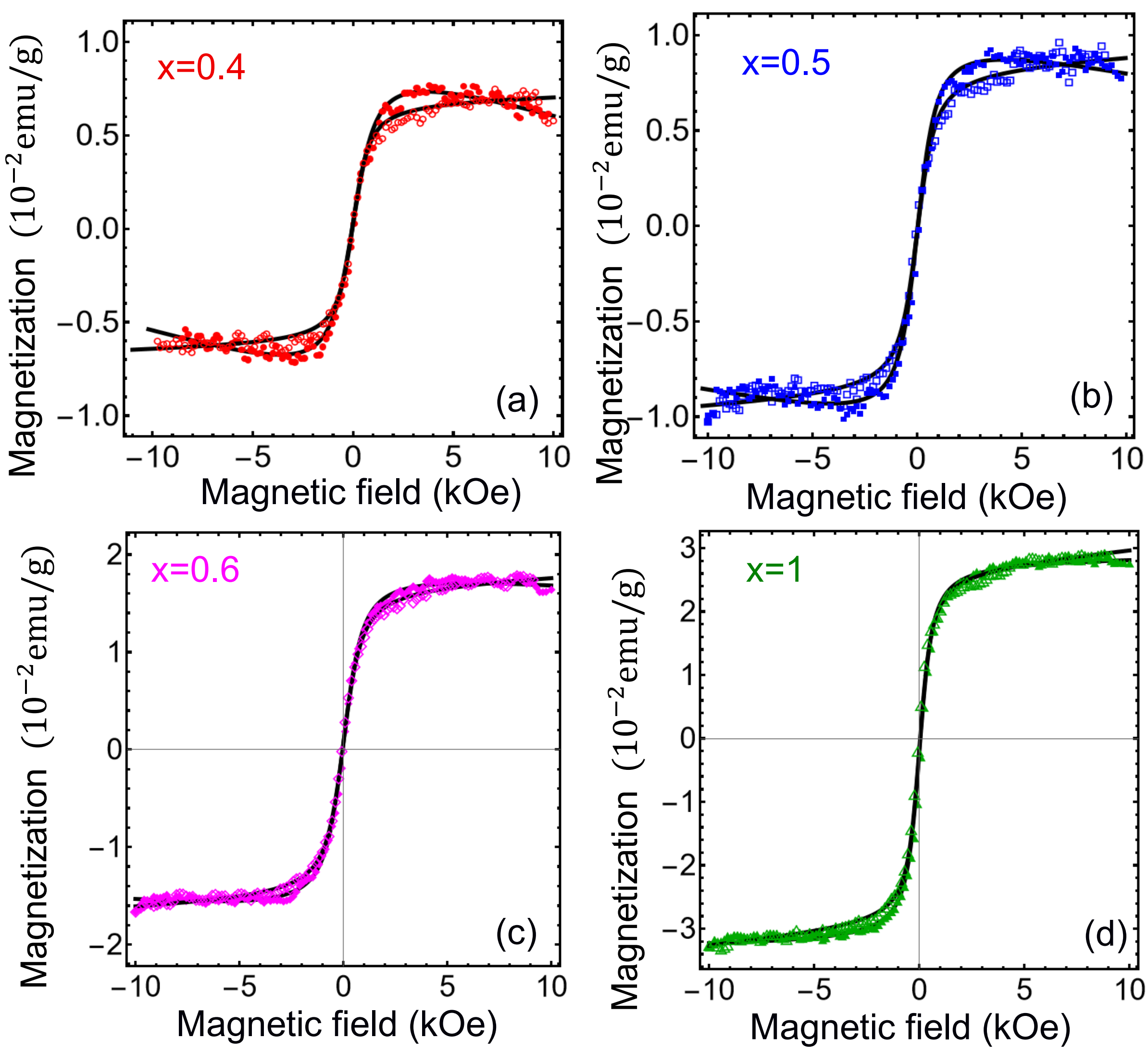

**FIGURE 6.** Dependences of the magnetization vs. the magnetic field applied to the $Hf_xZr_{1-x}O_{2-y}$ nanopowders with different hafnium content x=1 **(a)**, 0.6 **(b)**, 0.5 **(c)** and 0.4 **(d)**. Filled and empty symbols are experimentally measured values for the forward and backward direction of the magnetic field sweep, respectively. Black curves are calculated using expressions (3)-(4). The fitting parameters are listed in **Table I**, $T = 298$ K.

**Table I.** Fitting parameters for the magnetization-field dependence of $Hf_xZr_{1-x}O_{2-y}$ nanopowders

| Hf content x | field sweep direction | $\chi_m$ ($10^{-7}$ emu) | $M_S$ ($10^{-2}$ emu/g) | $\tilde{\mu}_s$ ($10^{-3}$ Oe$^{-1}$) | $\delta M$ ($10^{-2}$ emu/g) | $N_S^*$ ($10^{15}$ 1/g) |
|---|---|---|---|---|---|---|
| 0.4 | backward | 0.314 | 0.661 | 3.22 | 0.03 | 0.051 |
| | forward | -3.21 | 0.937 | 2.20 | 0.03 | 0.105 |
| 0.5 | backward | 0.703 | 0.868 | 3.06 | 0.03 | 0.070 |
| | forward | -2.18 | 1.084 | 2.83 | 0.03 | 0.095 |
| 0.6 | backward | 1.04 | 1.631 | 3.03 | 0.075 | 0.133 |
| | forward | -1.69 | 1.840 | 2.83 | 0.075 | 0.161 |
| 1.0 | backward | 3.37 | 2.837 | 4.52 | -0.150 | 0.155 |
| | forward | 0.322 | 3.058 | 4.21 | -0.210 | 0.180 |

*For the estimation we used that $N_S \cong M_S/\mu$, $\mu = \tilde{\mu} k_B T$, $k_B T$ =4.04530157×10$^{-14}$ erg at 298 K, and 1 emu of magnetic moment is equivalent to 1 erg/Oe.

In **Supplement S4** we estimated the relative concentration of paramagnetic ions in the $Hf_xZr_{1-x}O_{2-y}$ nanopowders, assuming that only $Hf(Zr)^{3+}$ ions (in complex with oxygen vacancies) contribute to saturated magnetization $M_S$. Results are listed in **Table S4**. Calculated fraction $\frac{N_{\mathrm{Hf(Zr)}^{3+}}}{N_{\mathrm{Hf(Zr)}}}$correlates with the area of defect-related Raman spectra for all "x" except for x = 0.35 – 0.4. The estimated value of $\frac{N_{\mathrm{Hf(Zr)}^{3+}}}{N_{\mathrm{Hf(Zr)}}} \sim 0.1$ % is too small to be determined by the XPS or EELS methods, but it is enough to produce SPM-like magnetization and high intensity EPR signal, because the sensitivity of EPR is $10^9$ spins/G even at room temperature. Note that the small values are related to the bulk of the nanoparticles and are calculated in the assumption that the dielectric/paramagnetic part of the response (proportional to $\chi_m H$) is excluded (see **Table 1**). The concentration of defects can exceed 5 % near the oxide surface in accordance with many experiments [57, 58, 59] and decreases exponentially far from the surface [60]. Assuming that the factor $\Delta R$ responsible for exponential decay of defect concentration in the nanoparticle shell is about one lattice constant (~0.5 nm) and the average radius $R$ of the nanoparticles is about 5 nm, the ratio $\frac{N_{\mathrm{Hf(Zr)}^{3+}}}{N_{\mathrm{Hf(Zr)}}}$ is about 0.25 % (at x = 0.4) and 1 % (at x = 1) in the nanoparticle shell.

### B. Measurements of the dielectric response and its analysis

To study the dielectric response of the $Hf_xZr_{1-x}O_{2-y}$ nanopowders we carried out capacitance measurements along with the charge transport measurements in the AC regime. The tableted powder samples were placed in a Teflon cylinder cell between two brass plungers, which create a uniaxial pressure of 2.5 MPa and serve as electrical contacts [41]. The samples had the disk shape with the diameter $D$ =4 mm and the thickness $d$ =(300 ± 20) μm ($d$ equals to the distance between the contacts).

Measurements of the capacitance and resistance of the pressed $Hf_xZr_{1-x}O_{2-y}$ nanopowders were carried out by the RLC-meter LCX200 ROHDE & SCHWARZ in the frequency range 4 Hz – 500 kHz. In result, a "colossal" dielectric response was observed [41]. In particular, the effective dielectric permittivity of the pressed $Hf_xZr_{1-x}O_{2-y}$ nanopowders reaches the values ~ $10^5$ – $10^6$ at low frequencies, at that the maximum of permittivity is diffuse and corresponds to the temperature range of 38 – 88°C (see Ref. [41] for details). The features are important for further analysis; they are highlighted in **Fig. 7.** Experimental data shown in the figure are borrowed from the data presented in Ref. [41].

Note that the permittivity increases with decreasing hafnium content "x" from 1 to 0.4, especially in the high-temperature wing (compare different curves in **Fig. 7(a)**). Along with this, the temperature of the dielectric permittivity maximum increases. For intermediate hafnium content (x~0.5) one can see a shoulder on the right wing of the permittivity and resistivity temperature dependences. All curves are quite close to each other and have similar shape in the left wing (before the minimum) of the resistivity temperature dependences, while they considerably differ in magnitude in the right wing (after the minimum).

The behavior can be explained by the gradual increase in the o-phase fraction with decrease in the hafnium content "x" due to the stabilizing role of zirconium atoms in the nanoparticles core. The increase of the permittivity is opposite to the magnetization decrease with decrease in "x" (see different magnetization curves in **Fig. 4(a)**), as well as it anti-correlates with the decrease in the intensity of Raman spectra with decrease in "x" (compare green, magenta and blue spectra in **Fig. 3**). The anti-correlation of the permittivity and magnetization can be explained by the bulk-type formation of dielectric response vs. the surface-type formation of the magnetic response. A colossal dielectric permittivity ($>10^5$) of the pressed $Hf_xZr_{1-x}O_{2-y}$ nanopowders is observed in a relatively low frequency range. The permittivity decreases monotonically from $10^5$ to the moderate values ~20 – 30 with increase in frequency from 4 Hz to 500 kHz (see different curves in **Fig. 7(b)** and Fig. 2 in Ref.[41]). The resistivity of the powder samples decreases from $10^9$ Ω·cm to $10^6$ Ω·cm with increase in frequency from 4 Hz to 500 kHz. The resistivity curves of the samples with higher content of zirconium tend to decrease again with further increase in temperature. The decrease occurs after passing the minimum and some increase (see red and blue curves with empty symbols in **Fig. 7(a))**. The changes of the resistivity temperature dependences with increase in the zirconium content "1-x" is important for understanding the DC resistivity temperature dependences. Further variation of resistivity at higher temperatures (up to 650 K) was carried out in the DC regime and presented in next subsection.

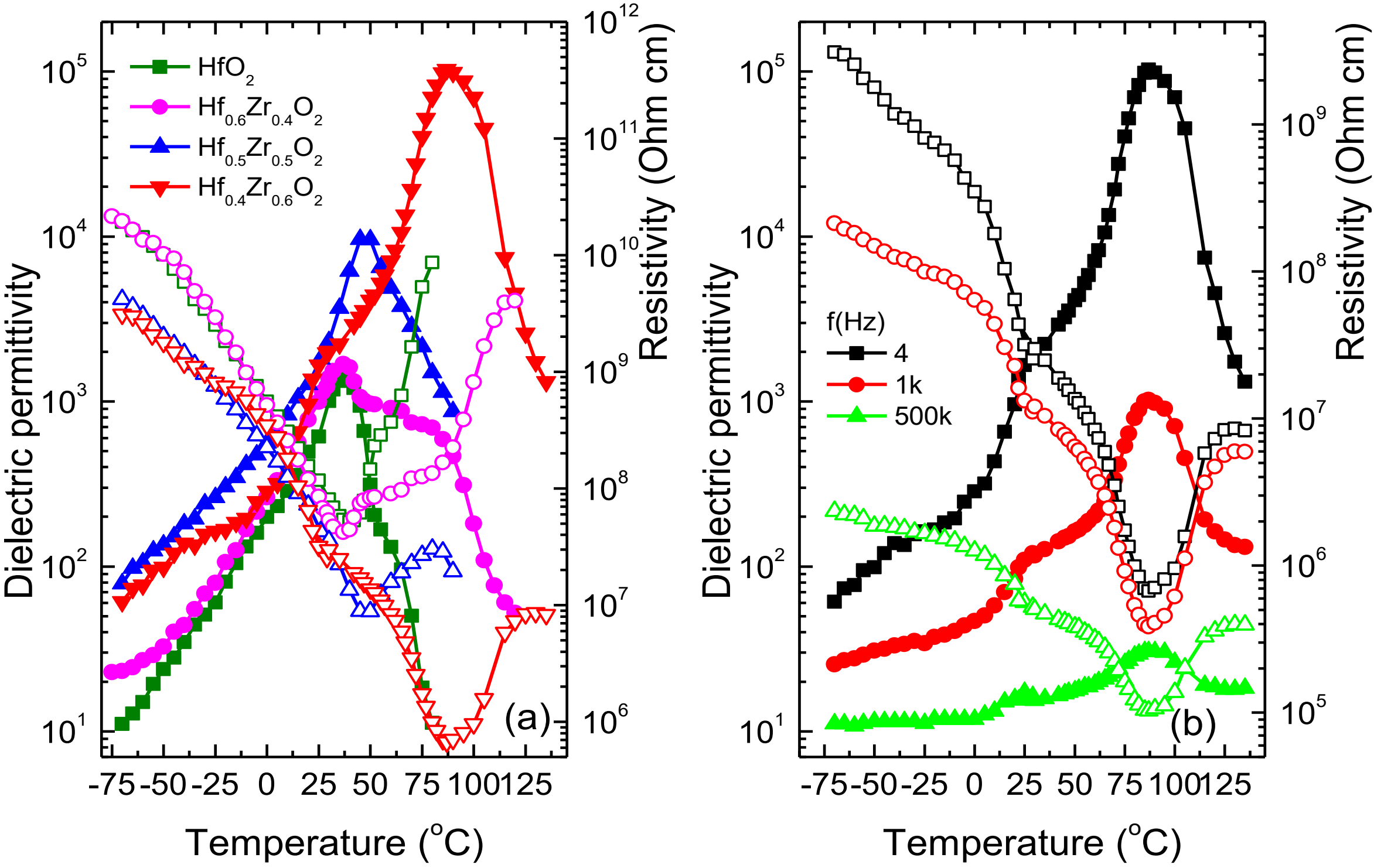

**FIGURE 7. (a)** Temperature dependences of the relative effective dielectric permittivity (filled symbols) and resistivity (empty symbols) of the pressed $Hf_xZr_{1-x}O_{2-y}$ nanopowders measured at the frequency 4 Hz. **(b)** Temperature dependences of the effective relative dielectric permittivity (filled symbols) and resistivity (empty symbols) of the pressed $Hf_{0.4}Zr_{0.6}O_{2-y}$ nanopowders measured at frequencies 4 Hz, 1 kHz to 500 kHz. The amplitude of the test sinusoidal voltage is 100 mV.

The strong decrease of the dielectric permittivity from the colossal to moderate values is accompanied by the increase in conductivity, which happens under the frequency increase. This behavior is typical for inhomogeneous ferroelectric-semiconductor materials, such as fine-grained ferroelectric ceramics and nanocomposites [61, 62, 63]. As a rule, the behavior of the dielectric response and losses is related to the interfacial barrier-layer capacitance (IBLC) effect [61], as well as to the inhomogeneous layers between the electrodes and the sample, known as the surface barrier layer capacitance (SBLC) effect [62]. In such cases, theoretical modeling of the nanopowders dielectric response can be performed using the effective medium approximation (EMA), namely the algebraic equation for the effective dielectric permittivity of the binary mixture (see details in Ref. [41]). Results of the theoretical modelling [41] show that the colossal dielectric permittivity observed at low frequencies can originate from the IBLC and/or SBLC effects in the orthorhombic cores of $Hf_xZr_{1-x}O_{2-y}$ nanoparticles.

Let us underline that the physical mechanism proposed for the explanation of the observed magnetic and dielectric properties of the $Hf_xZr_{1-x}O_{2-y}$ nanoparticles requires the presence of a large number of defects (e.g., either charge-neutral or ionized oxygen vacancies) near the surface of the nanoparticle. The presence of the defects in the $Hf_xZr_{1-x}O_{2-y}$ samples was confirmed by EDS, EELS and EPR methods, but their charge state, as well as the fraction of charged vacancies and its spatial redistribution, are unknown a priory. The charged oxygen vacancy with a high probability becomes neutral at the surface of $Hf_xZr_{1-x}O_{2-y}$ nanoparticle due to the complete screening by external free charges in the ambient conditions [15-17] and redox reactions [28]. Charged oxygen vacancies, as well as other charged defects, strongly influence the polar and electronic properties of nanosized hafnia-zirconia oxides [29], as well as participate in the charge transport (either as mobile charge defects or/and via the carrier hopping mechanism). Therefore, below we analyze the charge transport mechanisms of the $Hf_xZr_{1-x}O_{2-y}$ nanopowders with special attention to the ionic-type conduction that may be ascribed to the presence of oxygen vacancies and reflects the possible changes in their charge state.

## IV. DC CHARGE TRANSPORT MECHANISMS AND TRANSIENT PROCESSES IN $Hf_xZr_{1-x}O_{2-y}$ NANOPOWDERS

The measurements of charge transport properties of the $Hf_xZr_{1-x}O_{2-y}$ nanopowders in the DC regime were performed with the same samples as in the previous section. We carried out two kinds of experiments (for details see **Supplement S5** in **Supplementary Materials**).

In the first kind of experiment, we studied the transient processes in electric transport by applying the voltage pulse with duration of 60 s and different constant amplitude (from 1 to 60 V) to the samples. We measured the decay of electric current through the sample vs time at several temperatures in the temperature range around the minimum on the AC resistivity temperature curve. The sample was short-circuited for several minutes to guarantee the complete recovering of the initial state after each measurement of the current decay ~~time dependence~~.

In the second kind of experiment, the temperature dependence of DC resistivity was measured in the range from 180 to 650 K at a constant voltage drop ($U = 2$ V) across the sample. The time delay after each temperature change was sufficient to complete the transient current decay down to the steady-state value. In this kind of experiment the sample was not short-circuited after each temperature step.

The voltage power supply Instek PSP 603 was used as a source of a constant voltage in both kinds of experiments. The current was recorded by measuring the voltage drop across the series-connected 100 kΩ - load resistor by the Keithley 2000-SCAN precision multimeter.

We believe that these two kinds of experiments can help to determine the role of oxygen vacancies in electric charge transport. As it was shown earlier, the ferroelectric polarization of thin $Hf_{0.5}Zr_{0.5}O_2$ films often does not degrade but increases or "wake-up" after multiple electric field cycling [6, 22]. As a rule, the wake-up effect is conditioned by the slow redistribution and/or reordering of oxygen vacancies, which eventually can improve the polar and charge-accumulation properties of the films. Therefore, when the sample is not short-circuited after every temperature change, a slow drift of the vacancies (or/and the hopping of associated charged carriers) occurs, and the sample "wakes-up" towards better polar properties and a possible ferroelectric transition shifts to higher temperatures. This scenario corresponds to the second kind of experiment. When discharging or/and short-circuiting occurs after each temperature change, the polar ordering associated with the migration of vacancies cannot be formed and therefore useful wake-up effects are absent. This scenario corresponds to the first kind of experiment.

### IV.A. Transient processes

Our measurements reveal the following transient process in conduction of the pressed $Hf_{0.5}Zr_{0.5}O_{2-y}$ nanopowders. Namely, we observe a long-lasting decay of a current after applying a constant voltage to the sample, which may be related to the ionic-type conduction and charge accumulation process. These processes may be ascribed to migration and re-charging of oxygen vacancies, which are inherent to nanosized hafnia-zirconia (see e.g., the Ref. [28, 39-42] and references therein).

The current decay in the DC regime, unlike the rapid and strongly frequency-dependent resistive processes in AC regime, lasts up to several minutes in the $Hf_xZr_{1-x}O_{2-y}$ samples. For example, the waveforms of the current decay obtained at different amplitudes of the constant applied voltage in the range from 10 to 60 V are shown in **Fig. 8(a).** These curves are normalized by the maximum value at $t = 0$ for convenience. The non-normalized curves, shown in **Fig. 8(a)**, are well fitted by the two-exponential decay law:

$$J = J_0 + J_1 \exp\left(-\frac{t}{\tau_1}\right) + J_2 \exp\left(-\frac{t}{\tau_2}\right). \qquad (6)$$

This is illustrated in **Fig. 8(b)** for the fixed voltage amplitude $U$ =60 V. The fitting value $J_0$ =1.77 μA ~~at the voltage amplitude $U$ =60 V~~ corresponds to the steady-state DC conduction, whence the DC resistance of the sample is $3.4 \cdot 10^7$ Ω at 298 K. Decaying terms in Eq.(6) enable us to determine the accumulated charge $Q$ at each voltage drop and the capacitance $C$:

$$Q = J_1\tau_1 + J_2\tau_2, \qquad C = \frac{Q}{U}. \qquad (7)$$

The capacitance, corresponding to the long-lasting current decay in **Fig. 8(a)**, varies from 0.64 to 1.5 μF with the change in the applied voltage amplitude $U$ from 1 to 60 V, respectively. Using the ratio of sample sizes, $S/d$ =4.18, we calculated that the value of effective relative dielectric permittivity changes from $1.7 \cdot 10^6$ to $4 \cdot 10^6$. These values are colossal and inherent to the SPE state of the nanoparticles. Using the specific weight of $Hf_{0.5}Zr_{0.5}O_2$ (9.68 g/cm$^3$) and dimensions of the capacitor, we calculated that the specific capacitance $C/m$ ($m$ is the mass of the sample) varies from 17.5 to 41 μF/g in the mentioned range of applied voltage.

To explain the colossal values of the static permittivity within EMA model of a binary system, we assumed that the cores of $Hf_xZr_{1-x}O_{2-y}$ nanoparticles and their non-ferroelectric shells have different dielectric permittivity. The shell dielectric permittivity monotonically and relatively weakly depends on temperature but can be frequency dependent. Due to the ultra-small sizes, the cores are mostly in the SPE-like state induced by e.g., large flexo-electro-chemical strains [29]. Actually, the electric dipoles, which are the sources of polarization, are coupled with elastic defects (presumably oxygen vacancies) via the Vegard-type chemical strains, flexoelectric effect and electrostriction [29]. The flexo-electro-chemical coupling can induce the double potential wells in the order-disorder type thermodynamic functional of the nanoscale hafnia-zirconia oxides, being the driving force of the transition to the long-range ordered phase emerging with increase of the oxygen vacancies concentration [29].

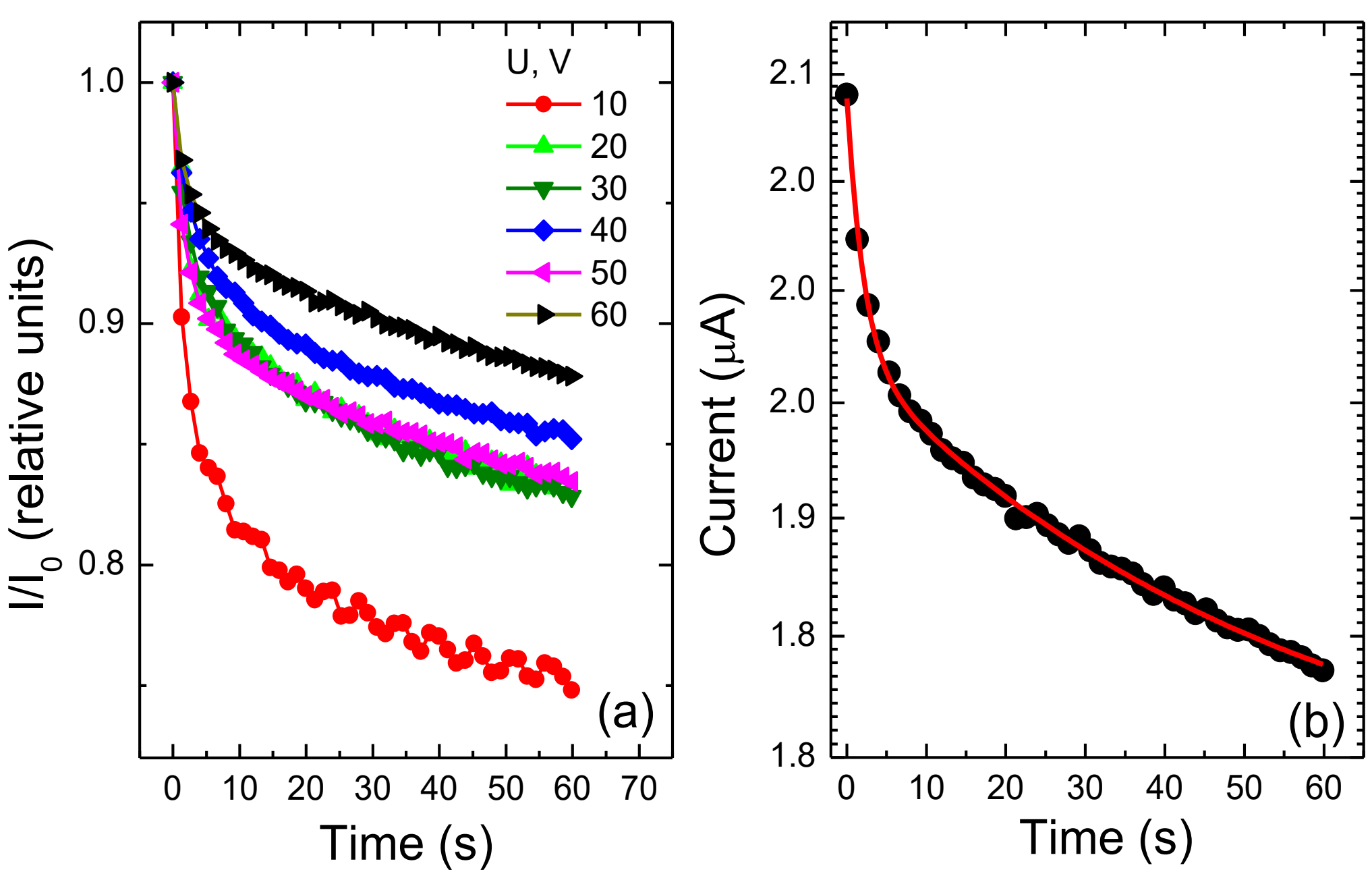

**FIGURE 8. (a)** The current vs time dependences at different amplitudes of applied voltage, which changes from 10 to 60 V. The curves are normalized to the maximum value at $t = 0$. **(b)** Fitting of the current decay curve, measured at voltage drop of 60 V, by the two-exponential decay law (6). Filled circles are experimental data for the pressed $Hf_{0.5}Zr_{0.5}O_{2-y}$ nanopowders, the red line is the fitting curve. The temperature $T$ =298 K.

The dependences of the decay constants $\tau_1$ and $\tau_2$ vs the applied voltage are shown in **Fig. 9(a)**. The transient process of the current decay completes totally in about 5 min (see **Fig. 9(b)**).

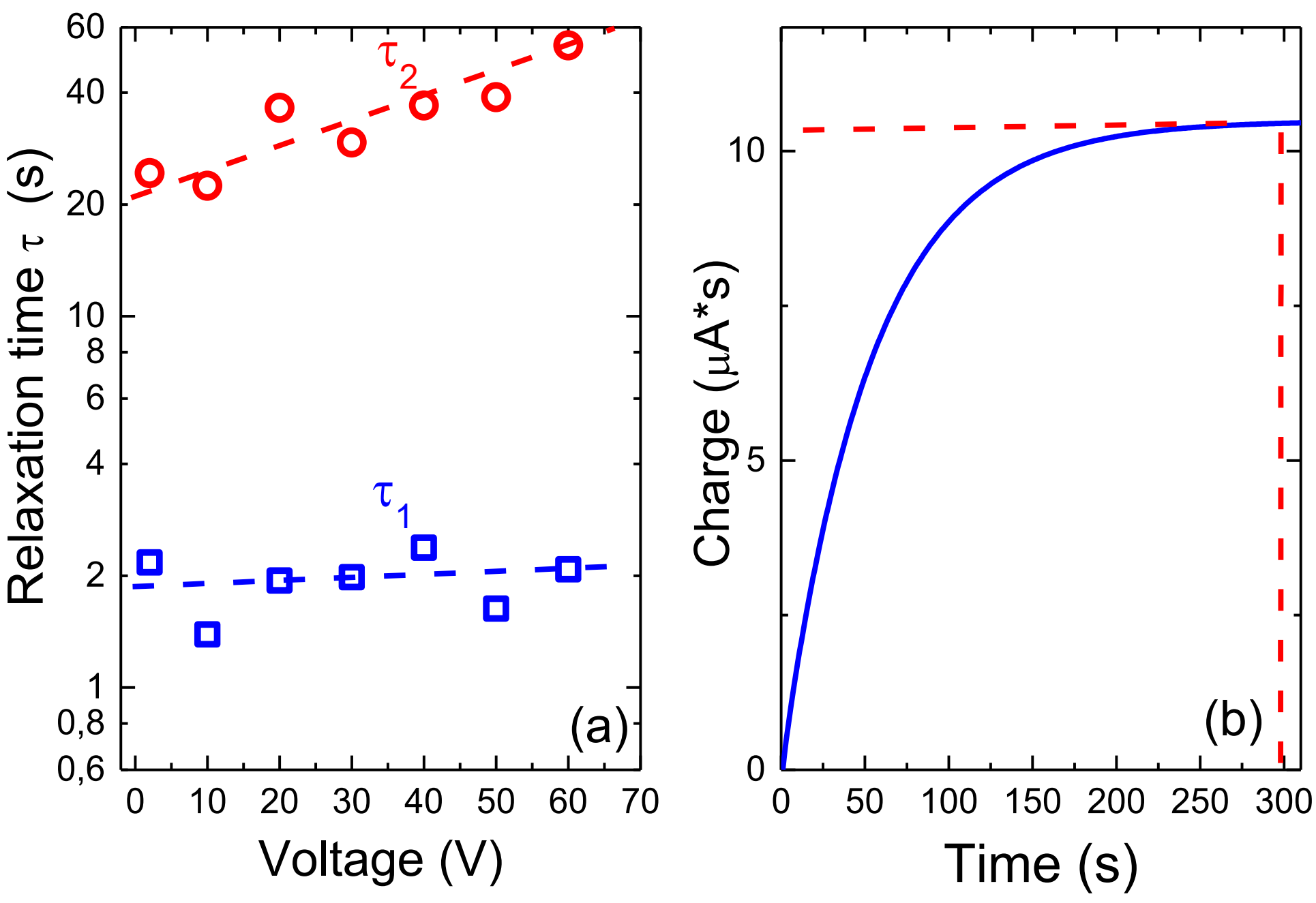

**FIGURE 9. (a)** The dependences of the time constants vs the voltage applied to the pressed $Hf_{0.5}Zr_{0.5}O_{2-y}$ nanopowders during the long-lasting current decay. **(b)** Charging of the sample during the transient process after applying voltage U=60 V across the pressed $Hf_{0.5}Zr_{0.5}O_{2-y}$ nanopowders; $T$ =298 K.

### IV.B. Temperature dependence of resistivity

The temperature dependences of the residual resistivity of $Hf_xZr_{1-x}O_{2-y}$ nanopowders measured in the DC regime at a constant voltage drop across the sample are shown in **Fig. 10**. Each point is recorded after a proper time delay at each temperature step required to complete the current decay. These measurements correspond to the second kind of experiment described at the beginning of Section IV. Corresponding temperature dependences measured in the AC regime (f = 4 Hz) are shown for comparison.

Depending on the resistance magnitude, the applied voltage varies from 2 to 20 V, which corresponds to the electric field strength of 5 – 50 V/cm. Measurements were performed in the temperature range of 200 – 650 K. The temperature range is limited by 650 K to avoid destroying the Teflon cell, and by the sample maximal resistivity level of $10^{10}$ Ω·cm. To avoid the possible influence of the strong heating on the results of measurements at low temperatures, the measurements were made first by cooling from 298 K to 200 K. Next the samples were heated back to the room temperature until the magnitude of the resistance was restored. Then the measurements were made under heating upwards in temperature.

The dependences shown in **Fig. 10** can be divided into two groups. The first group consists of pure $HfO_{2-y}$ and $Hf_xZr_{1-x}O_{2-y}$ samples with the lowest content of Zr. These dependences, measured both in DC and AC regimes, demonstrate a qualitatively similar behavior. They have a pronounced minimum, which shifts to lower temperatures in the AC regime. There is no tendency to decrease resistivity with an increase in temperature after passing the minimum. In contrast, the second group reveals completely different picture. After a small increase in resistivity on passing the minimum, the resistivity in the DC regime manifests a long decrease with an increase in temperature. Also, we observe a strong increase in resistivity with growing temperature above 570 K for the $Hf_{0.5}Zr_{0.5}O_{2-y}$ sample.

In fact, we observed a posistor-type effect in the $Hf_xZr_{1-x}O_{2-y}$ samples, which temperature range depends on the zirconium content. Note that a posistor effect in oxide semiconductors is well-known for the donor-doped grained or polycrystalline ferroelectric $BaTiO_3$ [64, 65, 66], where it is related with the transition from the paraelectric to the ferroelectric phase. The posistor effect can be explained by the temperature dependence of the dielectric permittivity with a strong maximum around the phase transition temperature within the Heywang model [64]. For the best of our knowledge, the observations of the posistor effect in the $HfO_2$-based materials are currently absent.

We observed only one minimum in the temperature dependence of resistivity in the $Hf_xZr_{1-x}O_{2-y}$ samples without or with small zirconium content. At the same time, a very pronounced minimum followed by the strong further increase of resistance (by more than an order of magnitude) is observed in the $Hf_{0.5}Zr_{0.5}O_{2-y}$ sample at 570 K (after a small minimum at 320 K). This can indicate the possible role of the oxygen vacancies in the behavior of charge transport. Note that the second minimum of resistivity is not observed in the sample with the highest zirconium content (x=0.4); possibly it is shifted to higher temperatures above the studied range.

We observe a very small slope of resistivity in the temperature range 300 – 350 K, determined in the DC regime in the $Hf_{0.5}Zr_{0.5}O_{2-y}$ sample, instead of a pronounced resistivity minimum observed in the AC regime. At the same time, one observes a pronounced maximum of the effective dielectric permittivity determined in the DC regime. The dielectric permittivity was calculated using Eq.(7) for the capacitance determined from the current decay curves measured at several points in the temperature range 300 – 350 K. So, one can conclude that the dielectric response shows similar behavior both in the AC and DC regimes in this temperature range.

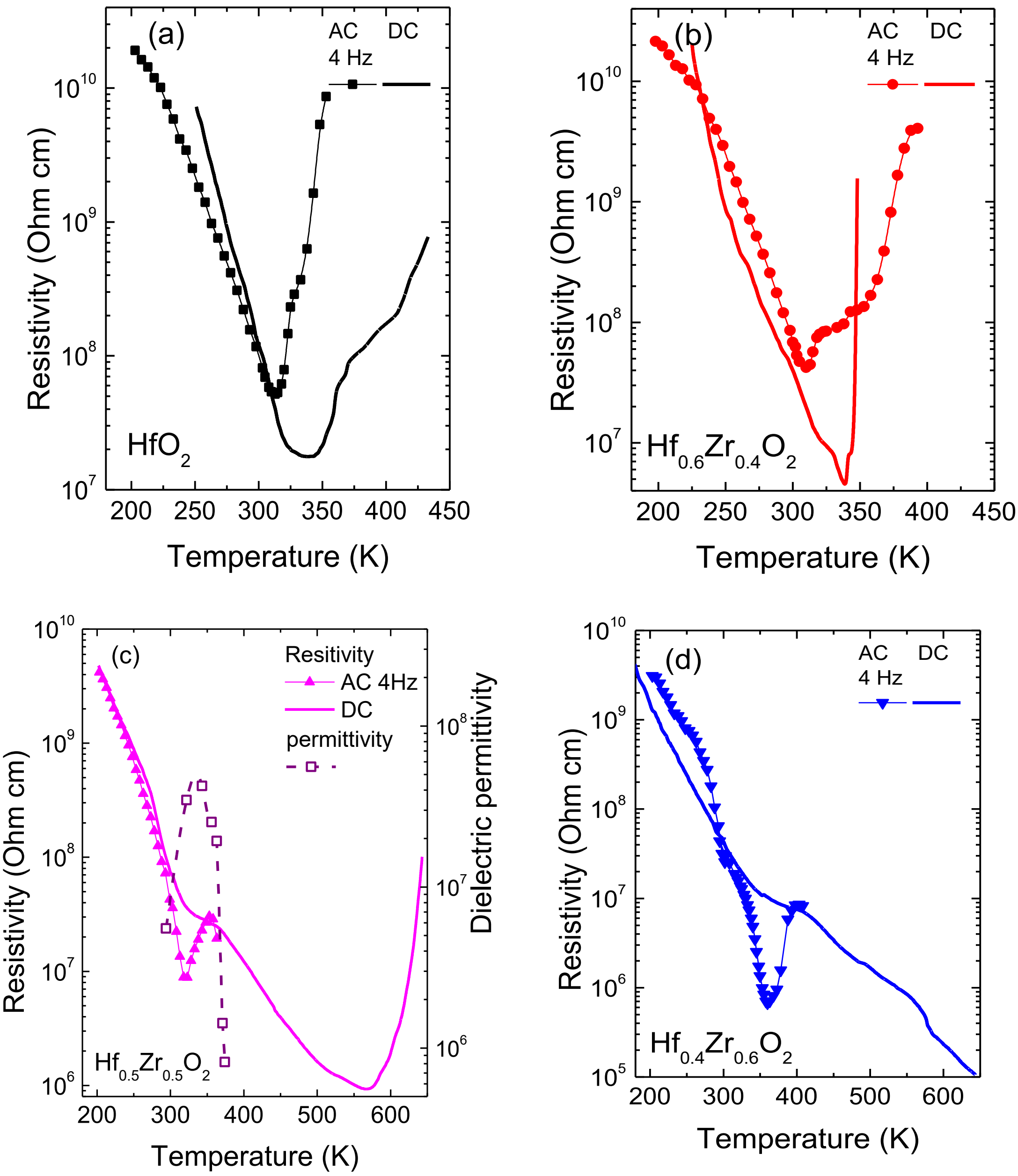

**FIGURE 10.** Temperature dependences of the resistivity of the $Hf_xZr_{1-x}O_{2-y}$ samples with x = 1 **(a)**, 0.6 **(b)**, 0.5 **(c)** and 0.4 **(d)**, measured in the DC mode (solid curves) and in the AC mode at 4 Hz (symbols). The temperature dependence of the effective dielectric permittivity corresponding to the DC regime is shown in part **(c)** by the empty rectangles connected by a dashed curve. The dependence is determined from the current decay curves measured at fixed temperature in the range 294 – 374 K.

Next, let us consider possible conduction mechanisms corresponding to the observed temperature dependences of the resistivity measured in the DC regime. It is well-known that the conduction in such materials is based on the model of hopping conduction, which has a semiconductor-like temperature dependence of resistivity. The fitting of the $Hf_{0.5}Zr_{0.5}O_{2-y}$ sample resistivity showed that it is well described in coordinates corresponding to the Mott law of the variable range hopping (VRH) conduction [67]:

$$R(T) \sim exp\left(\frac{T_0}{T}\right)^{1/4}. \quad (8)$$

The fitting of the $Hf_{0.5}Zr_{0.5}O_{2-y}$ sample resistivity in the Mott coordinates is shown in **Fig. 11.** The temperature dependence of the DC effective permittivity is also shown in the figure. It is seen that the resistivity temperature dependence very well obeys the Mott law except the temperature range of colossal dielectric permittivity (observed at 300 - 350 K) and the temperature range of the resistivity strong increase above 570 K. The first deviation from the Mott law can be explained in the Heywang model [68] considering the influence of the strong changes of dielectric permittivity on the conduction in ferroelectric-semiconductors. This model can be used also to explain appearance of the posistor effect.

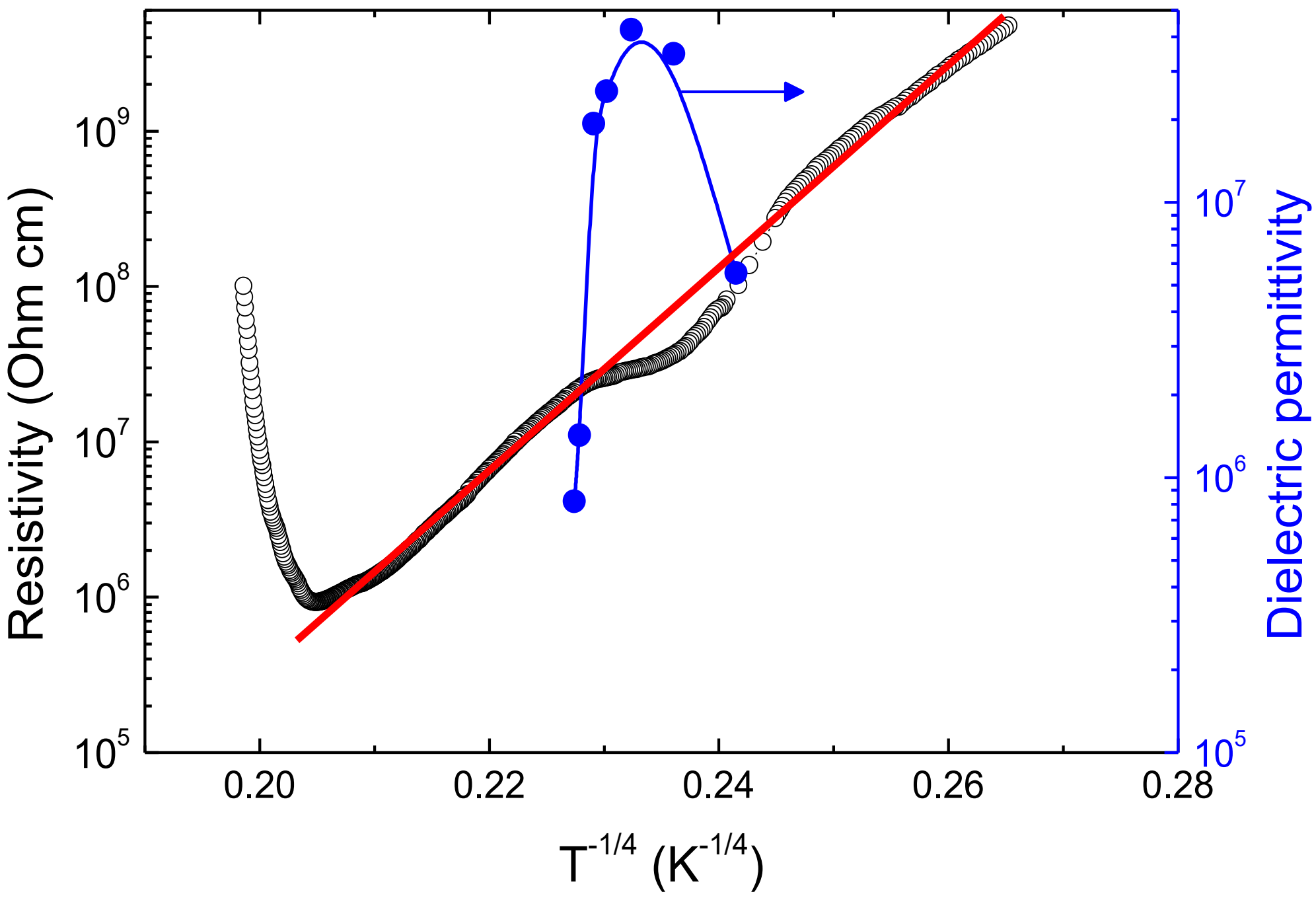

**FIGURE 11.** Temperature dependences of resistivity and effective dielectric permittivity of the $Hf_{0.5}Zr_{0.5}O_{2-y}$ sample in the coordinates corresponding to the Mott law of the VRH.

From fitting the dependence $R(T)$ for $Hf_{0.5}Zr_{0.5}O_{2-y}$ sample by Eq.(8) one obtains that $T_0 = 5 \cdot 10^8$ K ($T_0^{1/4}$=150 $K^{1/4}$), which is quite a large magnitude, though remains with reasonable limits for semiconducting materials. At the same time the dependence $R(T)$ is also well fitted in the Arrhenius coordinates (**Fig. 12**), where it has two activation parts with different activation energies, 327 and 154 meV, separated by the region 300 – 350 K, where the effective dielectric permittivity reaches colossal values in the maximum.

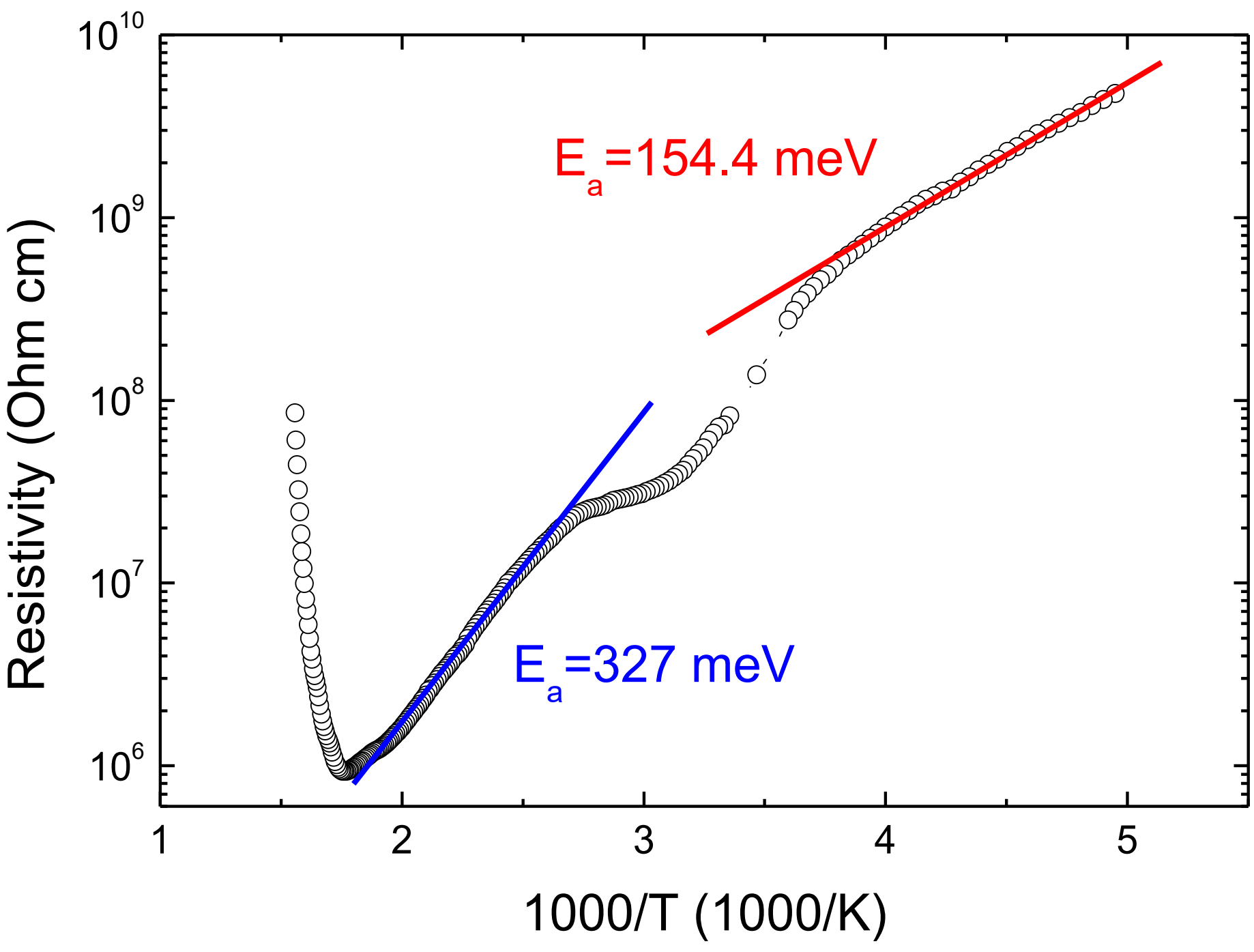

**FIGURE 12.** Temperature dependences of the $Hf_{0.5}Zr_{0.5}O_{2-y}$ sample resistivity in the Arrhenius coordinates.

The DC conduction in the $Hf_{0.5}Zr_{0.5}O_{2-y}$ is strongly determined by the ionic as well as electron transport, which results both in charge/discharge processes and leakage on switching on/off the applied voltage. The electron charge may leak into contacts while ionic charge may move only through the sample. The leakage currents are quite small in the pressed $Hf_{0.5}Zr_{0.5}O_{2-y}$ nanopowders and determined by the phonon-assisted hopping conduction between oxygen vacancies [69, 70, 71]. They depend substantially also on the whole defect state and doping. As follows from recent studies [41], it may depend also, as a conduction mechanism, on the dielectric permittivity, which in turn may depend strongly on the material structure, grain size, and temperature.

## V. CONCLUSIONS

We studied magnetic properties and electric transport mechanisms of the powder samples consisting of small (8 – 10 nm) oxygen-deficient $Hf_xZr_{1-x}O_{2-y}$ nanoparticles (x = 1 – 0.4) prepared by the solid-state organonitrate synthesis. We revealed that the static magnetization curves have a pronounced SPM-like behavior, and the value of saturated magnetization $M_s$ decreases monotonically with decrease in hafnium content “x” from 1 to 0.4. This tendency means the gradual decrease of the magnetic defect concentration with decrease in “x”. The influence of zirconium doping on $M_s$ saturates for $x < 0.5$, indicating the possible role of magnetic defects accumulated near the surface.

In this work we observed also that the magnitude of static dielectric permittivity changes from $1.7 \cdot 10^6$ to $4 \cdot 10^7$ for the $Hf_{0.5}Zr_{0.5}O_{2-y}$ nanopowders. Such values are colossal and inherent to the SPE state. Results of theoretical modelling, based on the EMA approach, show that the colossal dielectric permittivity observed at low frequencies can originate from the IBLC and/or SBLC effects in the orthorhombic cores of the $Hf_xZr_{1-x}O_{2-y}$ nanoparticles. Due to the ultra-small sizes, the cores are mostly in the SPE-like state induced by the flexo-electro-chemical strains. We conclude that the flexo-electro-chemical coupling may be the driving force of the $Hf_xZr_{1-x}O_{2-y}$ nanoparticles transition to the long-range ordered state emerging with increase of defect concentration. The SPE cores contribute to the colossal dielectric response, while the ultra-thin magnetic shells are responsible for the SPM response of the nanoparticles.

The EPR spectra of $Hf_xZr_{1-x}O_{2-y}$ nanopowders reveal the presence of paramagnetic defect centers, which may be hafnium and/or zirconium ions that trapped an electron near an oxygen vacancy and changed their valence state from +4 to the paramagnetic +3 state. The shape of the Raman spectra indicates the decisive role of surface defects, presumably oxygen vacancies, for all studied x = 1 – 0.4. At the same time the EDS and EELS analysis did not reveal any noticeable concentration of magnetic impurities in the $Hf_xZr_{1-x}O_{2-y}$ nanopowders.

Performed measurements allow to determine the conduction mechanisms in the pressed $Hf_xZr_{1-x}O_{2-y}$ nanopowders. In particular, we observed a long-lasting decay of the DC current after applying a constant voltage to a sample, which is related to ionic-type conduction and some charge accumulation process. The ionic-type conduction and the charge accumulation can be ascribed to the migration and re-charging of oxygen vacancies, which are accumulated in the studied nanoparticles.

The temperature dependences of the $Hf_xZr_{1-x}O_{2-y}$ resistivity demonstrate a posistor-like behavior in the temperature range that depends on the hafnium content "x". At that the temperature dependences of resistance in the range, where they possess the negative temperature coefficient of resistivity, obey the laws corresponding to the VRH conduction mechanisms.

The obtained results open the way for creation of innovative silicon-compatible ferroics – $Hf_xZr_{1-x}O_{2-y}$ nanoparticles with SPM and SPE properties, which may become indispensable ultra-high k nanomaterials for advanced FETs and electronic logic elements.

## Supplementary Materials

Supporting Information containing preparation details of the oxygen-deficient $Hf_xZr_{1-x}O_2$ nanoparticles and their XRD characterization (**Supplement S1**), EDS studies (**Supplement S2**) Raman analysis (**Supplement S3**), and calculation details of magnetization (**Supplement S4**), measurements of resistivity (**Supplement S5**), are given in Supplementary Materials.

## Authors' contribution

The research idea and results interpretation belong to A.N.M. and V.V.V. O.S.P., and D.O.S. conducted electrical measurements, V.V.V. interpreted and fitted the results. E.A.E. and A.N.M.

performed analytical calculations and fitting the magnetic measurements results. A.V.B. performed magnetic measurements and O.S.P. performed data processing. V.V.L. and Y.O.Z. measured EPR spectra and analyzed the results. O.V.L. and V.N.P. sintered the $Hf_xZr_{1-x}O_2$ nanopowders. L.D. performed TEM, EDS and EELS studies. V.I.S. performed SEM studies. M.V.K. performed XRD studies. A.D.Y. and O.M.F. measured Raman spectra and analyzed the results. A.N.M. and V.V.V. wrote the manuscript draft. All co-authors discussed and analyzed the results, and corresponding authors made all improvements in the manuscript.

## Acknowledgments

Authors are very grateful to the Reviewers for useful remarks and stimulating discussions. O.S.P. D.O.S., A.D.Y. and O.M.F. and A.N.M. acknowledge the support the National Research Foundation of Ukraine (grant N 2023.03/0132 “Manyfold-degenerated metastable states of spontaneous polarization in nanoferroics: theory, experiment and perspectives for digital nanoelectronics”). The work of E.A.E. and Y.O.Z is funded by the National Research Foundation of Ukraine (grant N 2023.03/0127 “Silicon-compatible ferroelectric nanocomposites for electronics and sensors”). The work of A.V.B. is funded by the NAS of Ukraine, grant No. 07/01-2025(6) "Nano-sized multiferroics with improved magnetocaloric properties". L.D. acknowledges support from the Knut and Alice Wallenberg Foundation (grant no. 2018.0237) for TEM research. V.V.V. acknowledge the Target Program of the National Academy of Sciences of Ukraine, Project No. 5.8/25-П “Energy-saving and environmentally friendly nanoscale ferroics for the development of sensorics, nanoelectronics and spintronics”. A.N.M. also acknowledges the Horizon Europe Framework Programme (HORIZON-TMA-MSCA-SE), project № 101131229, Piezoelectricity in 2D-materials: materials, modeling, and applications (PIEZO 2D). Results of theoretical modelling were visualized in Mathematica 14.0 [72].

## Data availability

The data is available at a reasonable request to the authors.

## References Supplementary Materials

### SUPPLEMENT S1. Preparation of the oxygen-deficient $Hf_xZr_{1-x}O_2$ nanoparticles and their XRD characterization

The preparation of the $Hf_xZr_{1-x}O_{2-y}$ nanopowders is listed in **Table SI** and described in detail in Ref. [73]. The XRD analysis was performed using the XRD-6000 diffractometer with Cu-Kα1 emission ($2\theta$ = 15 - 70°); and the database of the International Committee for Powder Diffraction Standards (JCPDS PDF-2) was used for identification of the crystallographic phases. In the result, we revealed the coexistence of the m-phase (space group P21/c, the content varies from 13 to 5 wt/ %) and three inseparable o-phases (space groups Pbca, Pbcm and ferroelectric Pca21, the content varies from 87 to 96 wt. %) in the nanopowder samples (see **Table S1**). Corresponding diffractograms are given in **Figure S1.**

**Table S1.** Characteristics of the $Hf_xZr_{1-x}O_{2-y}$ nanopowders. Adapted from Ref. [41].

| Sample chemical composition | Characteristics of the samples from XRD | | | | |
|---|---|---|---|---|---|
| | Lattice parameters (nm)* | | phase (wt. %), ** CSR size (nm) | | Preparation details |
| | o-phases | m-phase | o-phases | m-phase | |
| $HfO_{2-y}$ | $a$ = 1.0118<br>$b$ = 0.5202<br>$c$ = 0.5122 | $a$ = 0.5125<br>$b$ = 0.5158<br>$c$ = 0.5305 | 87 %<br><br>8 nm | 13 %<br><br>10 nm | decomposition of nitrates at 500°C, 6 hours in CO + $CO_2$ ambient |
| $Hf_{0.6}Zr_{0.4}O_{2-y}$ | $a$ = 10.108<br>$b$ = 0.5128<br>$c$ = 0.5160 | $a$ = 0.5148<br>$b$ = 0.5167<br>$c$ = 0.5331 | 95 %<br><br>8.5 nm | 5 %<br><br>11.5 nm | decomposition of nitrates at 600°C, 2 hours in CO + $CO_2$ ambient |
| $Hf_{0.5}Zr_{0.5}O_{2-y}$ | $a$ = 1.0062<br>$b$ = 0.5128<br>$c$ = 0.5160 | $a$ = 0.5146<br>$b$ = 0.5163<br>$c$ = 0.5328 | 96 %<br><br>10 nm | 4 %<br><br>10 nm | decomposition of nitrates at 600°C, 2 hours in CO + $CO_2$ ambient |
| $Hf_{0.4}Zr_{0.6}O_{2-y}$ | $a$ = 1.0109<br>$b$ = 0.5128<br>$c$ = 0.5160 | $a$ = 0.5130<br>$b$ = 0.5208<br>$c$ = 0.5307 | 89 %<br><br>9 nm | 11 %<br><br>7 nm | decomposition of nitrates at 600°C, 2 hours in CO + $CO_2$ ambient |

*The monoclinic angle 99.2 degrees for the m-phase. This angle was fixed for all samples during the refinement process.

**The size of coherent scattering regions (CSR) differs from the particle sizes determined directly (e.g., from TEM), however the average sizes well correlate with the size of CSR for the o-phase.

***As it was shown by Fujimoto et al. [74], the decomposition of the XRD spectra of $Hf_{0.5}Zr_{0.5}O_{2-y}$ nanoparticles can be done using the "m + o" basis and the "m + t" basis, where "t" means the tetragonal phase. However, the decomposition of the XRD spectra using the full "m + t + o" basis, appeared doubtful due to the proximity of the o- and t-phases spectral lines. In the work [41] we start fitting the XRD spectrum using "m + o" basic and reach a good convergency with different lattice constants $a$, $b$ and $c$, listed in **Table S1**. The condition $a = b$ should be valid in the t-phase. When one start from the "m + t + o" basic in the considered case of the $Hf_xZr_{1-x}O_{2-y}$ nanopowders, the axes do not become equal at the end of the fitting time. We have tested this result for all samples, which confirmed that we can trust in the "o + m" case [41], probably due to the large amount of oxygen vacancies stabilizing the o-phases.

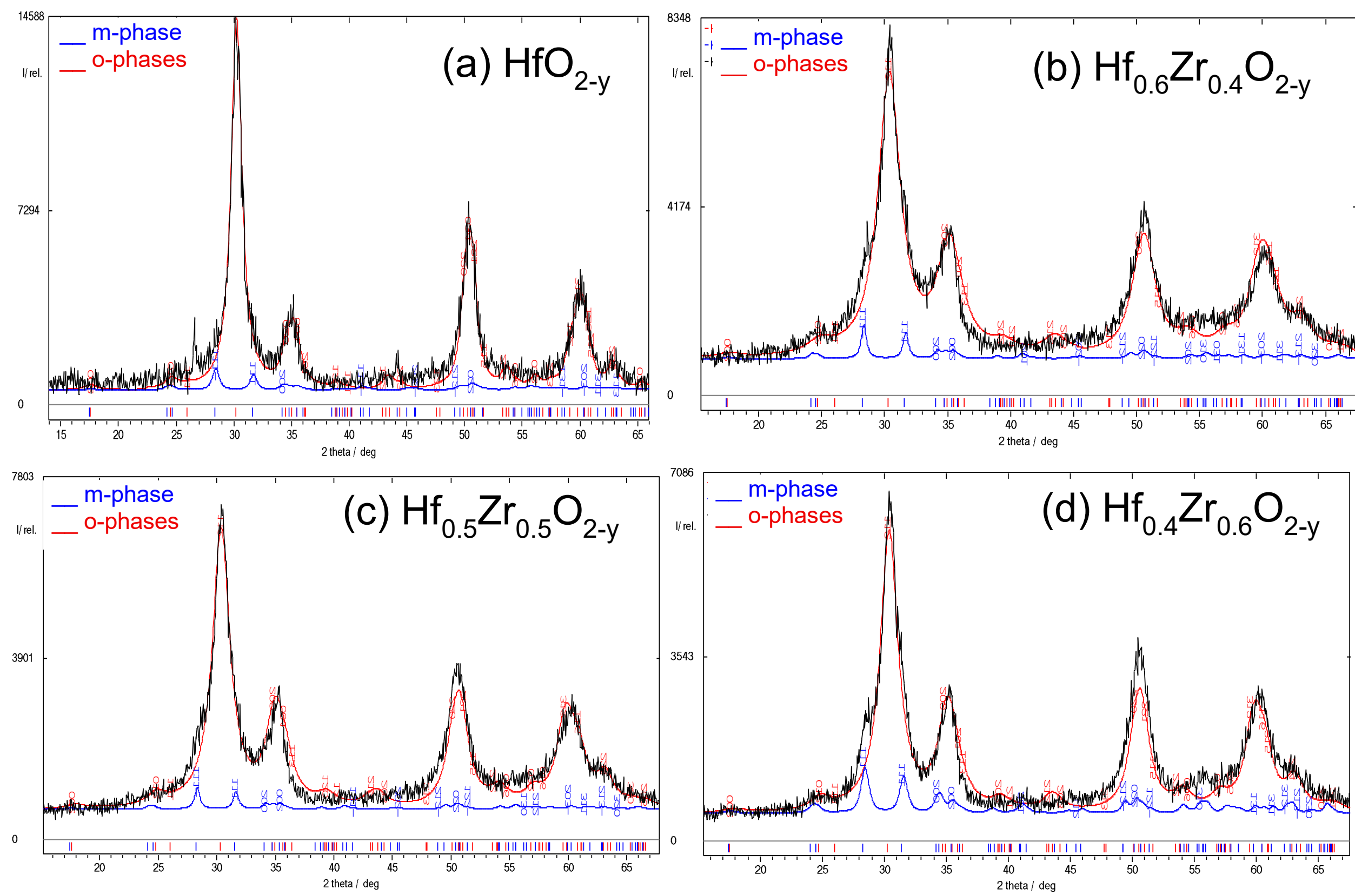

**FIGURE S1.** XRD spectra of the $Hf_xZr_{1-x}O_{2-y}$ nanopowders. Adapted from Ref. [41].

## SUPPLEMENT S2. Analysis of chemical composition by EDS and EELS

### S2.A. Energy-dispersive X-ray spectroscopy analysis of $Hf_{1-x}Zr_xO_2$ nanopowders

The Energy-dispersive X-ray spectroscopy (EDS) was employed to determine the chemical composition of the $Hf_xZr_{1-x}O_{2-y}$ samples. The EDS spectra were acquired using the JEOL JEM-2100F transmission electron microscope equipped with an EDS detector, operating at an accelerating voltage of 200 kV.

Quantitative analysis was performed using a standard quantification procedure with Cliff–Lorimer correction. For zirconium and hafnium, the Zr Kα and Hf Lα characteristic lines were used, respectively, while the O Kα line was used for oxygen. Due to the well-known limitations of EDS quantification of light elements, particularly oxygen, the oxygen content was considered semi-quantitative. Due to strong absorption effects and thickness-dependent oxygen Kα emission, the oxygen concentration obtained by EDS should be treated with caution. Therefore, the discussion focuses primarily on the Hf/Zr cation ratio, which is reliably determined under applied experimental conditions.

The elemental ratios of hafnium and zirconium were determined directly from the integrated peak intensities of the corresponding characteristic lines. The measured atomic percentages were normalized to the total metal content (Hf + Zr = 1) to obtain the effective metal-site composition of the oxide, expressed as $Hf_xZr_{1-x}O_{2-y}$.

Multiple spectra were acquired from different regions of each sample to ensure reproducibility. The estimated uncertainty of the Hf/Zr atomic ratio determined by TEM-EDS is approximately ±0.5 at.%, while the corresponding normalized metal-site fraction uncertainty is about ±0.005.

Representative TEM-EDS spectra are shown in **Fig. S2**. The relative concentrations of hafnium and zirconium in the oxide samples were estimated from the intensities of the characteristic Hf and Zr lines observed in the EDS spectra. The comparison of these line intensities allows a reliable evaluation of the Hf/Zr cation ratio in the Zr–Hf oxide system.

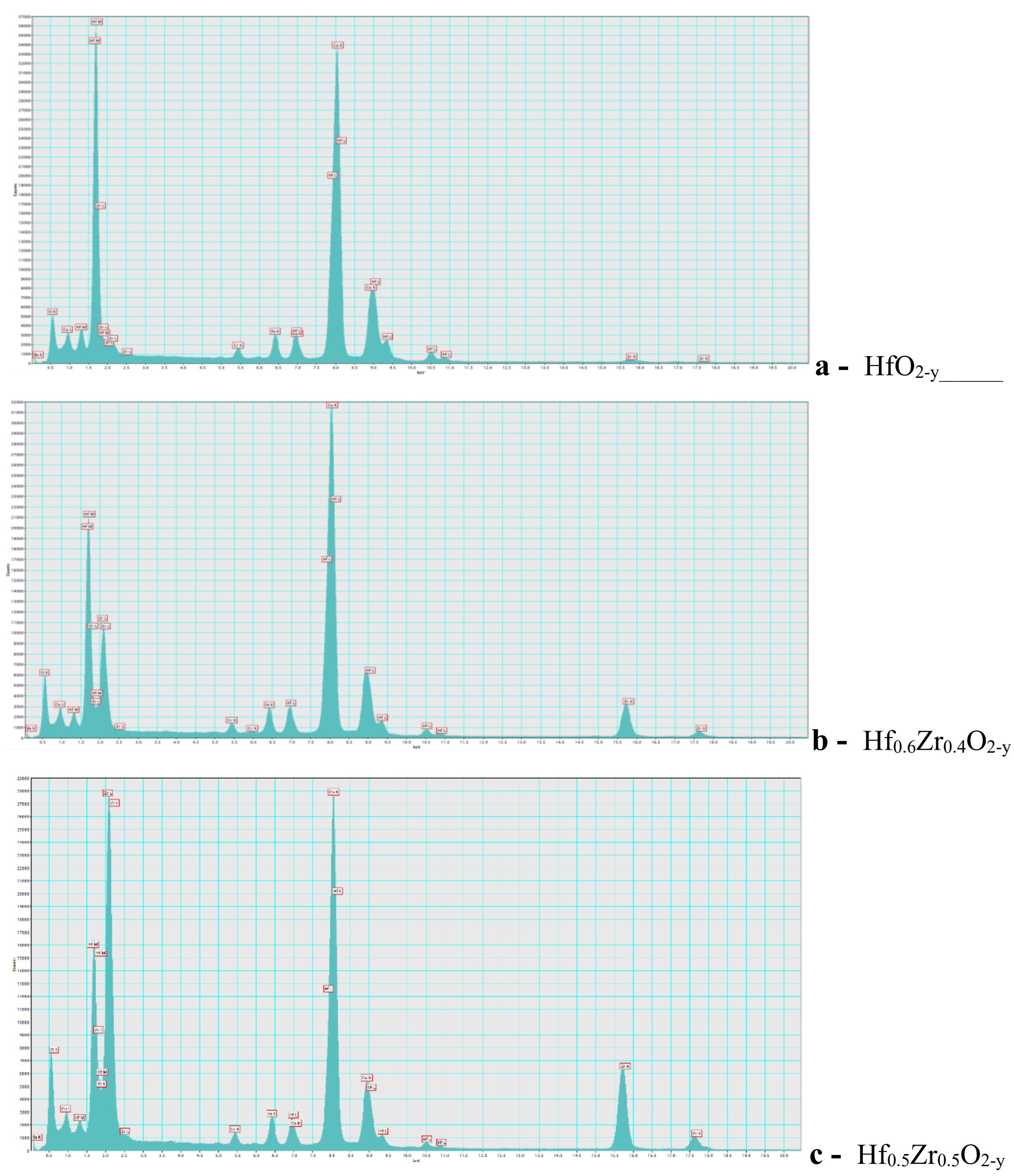

**a -** $HfO_{2-y}$

**b -** $Hf_{0.6}Zr_{0.4}O_{2-y}$

**c -** $Hf_{0.5}Zr_{0.5}O_{2-y}$

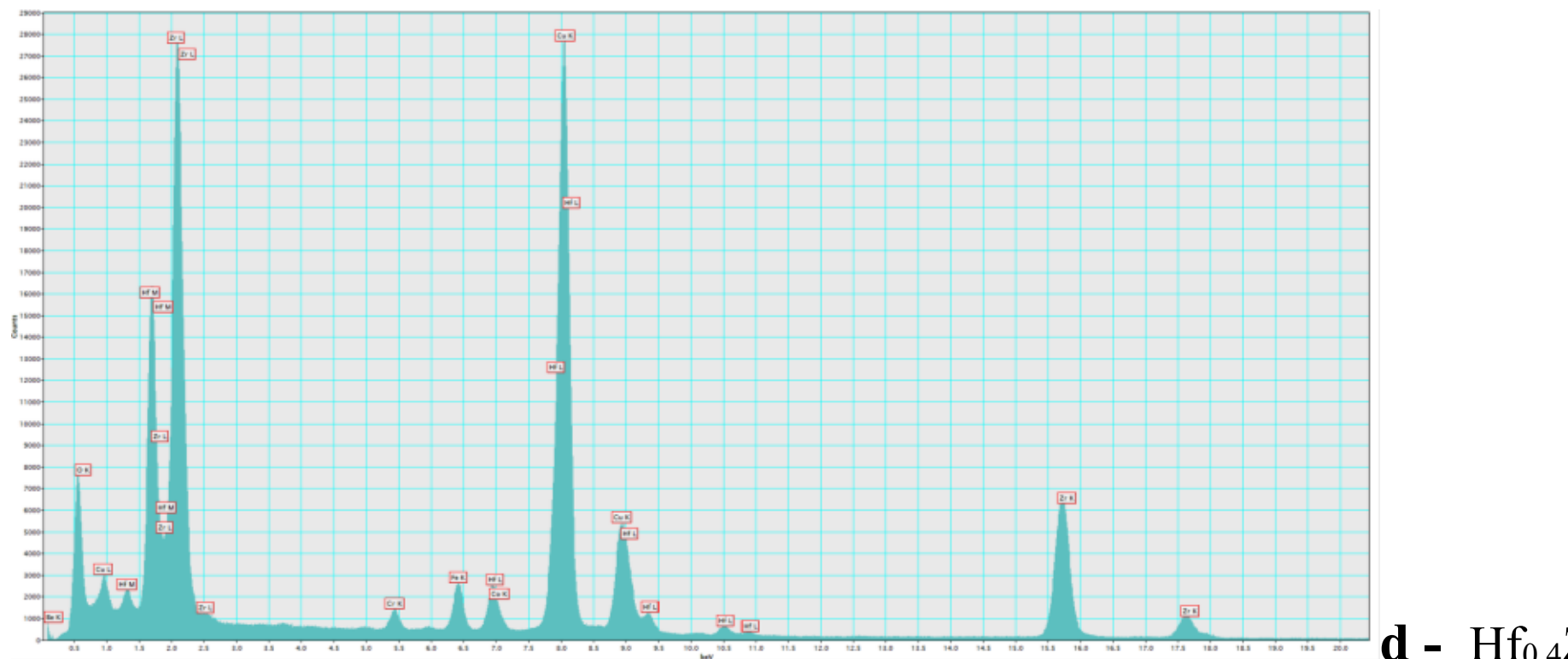

**d -** $Hf_{0.4}Zr_{0.6}O_{2-y}$

**FIGURE S2**. Typical EDS spectra for the $Hf_xZr_{1-x}O_{2-y}$ nanopowders: $HfO_{2-y}$ **(a),** $Hf_{0.6}Zr_{0.4}O_{2-y}$ **(b),** $Hf_{0.5}Zr_{0.5}O_{2-y}$ **(c),** and $Hf_{0.4}Zr_{0.6}O_{2-y}$ **(d).**

Results of TEM-EDS measurements and derived compositions of the $Hf_xZr_{1-x}O_{2-y}$ nanopowders are given in **Table S2**.

**Table S2.** Nominal and EDS-measured chemical composition of the $Hf_xZr_{1-x}O_{2-y}$ nanopowders

| **Nominal composition** | **EDS-measured Hf/Zr ratio, at % ±0.5** | | **Derived composition, ±0.005** |
|---|---|---|---|
| | Hf | Zr | |
| $HfO_{2-y}$ | 89.0 | 11.0 | $Hf_{0.89}Zr_{0.11}O_{2-y}$ |
| $Hf_{0.6}Zr_{0.4}O_{2-y}$ | 65.1 | 34.9 | $Hf_{0.65}Zr_{0.35}O_{2-y}$ |
| $Hf_{0.5}Zr_{0.5}O_{2-y}$ | 54.7 | 45.3 | $Hf_{0.55}Zr_{0.45}O_{2-y}$ |
| $Hf_{0.4}Zr_{0.6}O_{2-y}$ | 35.4 | 64.6 | $Hf_{0.35}Zr_{0.65}O_{2-y}$ |

In addition to the signals originating from the constituent elements of the samples, the EDS spectra also contain characteristic lines of copper and several other elements, which do not belong to the studied oxides. The presence of Cu peaks is attributed to the use of copper TEM grids for the sample support. Furthermore, weak signals from iron and cobalt are occasionally detected, which originate from the interaction of the electron beam with microscope components and surrounding structural parts. These contributions were excluded from the quantification procedure and do not affect the determination of the Hf/Zr ratio.

### S2.B. Electron Energy-Loss Spectroscopy (EELS) analysis of $Hf_{1-x}Zr_xO_2$ nanopowders

The Electron Energy Loss Near Edge structures (ELNES) originates from the electron transition from a core orbital to unoccupied bands. The Electron Energy-Loss Spectroscopy (EELS) analysis of $Hf_{1-x}Zr_xO_2$ nanopowders are discussed below. The EELS gives volumetric results, which are analogous to the XPS surface-related results (see comparison at https://eels.info/why-eels/overview).

**Figure S3(a)** shows a typical O K-edge EELS spectrum of $Hf_{1-x}Zr_xO_{2-y}$ nanoparticles recorded in the TEM mode (on example of the $Hf_{0.35}Zr_{0.65}O_{2-y}$ sample). The O K edge arises from transitions of O 1s core electrons into unoccupied O 2p states hybridized with the Hf 5d and Zr 4d orbitals and therefore probes the unoccupied metal *d*-derived electronic states. The spectrum exhibits four dominant features labeled as A, B, C and D.

To clarify the origin of the fine structure, the O K-edge spectrum was deconvoluted, as shown in **Fig. S3(b).** The deconvolution reveals contributions associated with the O 2p states hybridized with metal $t_{2g}$ and $e_g$ orbitals, as well as a low-energy pre-edge component preceding the main $t_{2g}$ peak. The peak A (located at ~532 eV) is assigned to the transitions into O 2p – $t_{2g}$ hybridized states, while the peak B (located at ~535.5 eV) at higher energy corresponds to the O 2p – $e_g$ hybridized states. These features reflect the octahedral crystal-field splitting of the metal *d*-orbitals. The energy separation between the two peaks defines the effective crystal-field splitting $\Delta_{oct}$ (splitting of the *d*-orbital of Hf and Zr due to octahedral ligand field). Compared to stoichiometric $ZrO_2$ and $HfO_2$, the $t_{2g}$-related peak A in $Hf_{1-x}Zr_xO_{2-y}$ has a significantly reduced intensity [75].

In addition, a "pre-edge" (or shoulder) appears on the low-energy side of the peak A at approximately 529–531 eV. The pre-edge is indicated by the red rectangle in **Fig. S3(b)**. This low-energy shoulder is attributed to oxygen vacancy–induced defect states [76, 77, 78]. Oxygen vacancies introduce excess electrons that partially occupy metal *d* states, reducing the number of unoccupied $t_{2g}$ states and, consequently, the intensity of the $t_{2g}$-related peak. The defect-related states are located below the conduction-band minimal energy and give rise to the transitions from O 1s to the localized donor-like states.

The extracted crystal-field splitting for the $Hf_{1-x}Zr_xO_{2-y}$ is $\Delta_{oct} \approx 3.5$ eV, which is smaller than the reported values for $ZrO_2$ (~4.0 eV) and $HfO_2$ (~5.0 eV) (see e.g., [75]). The field reduction indicates a modified metal-oxygen bonding environment in the mixed oxide system. The onset energy $E_0$ of the O K edge, defined as the inflection point of the rising edge, is located at approximately 530.5 eV. This value lies below the values of pure $ZrO_2$ and $HfO_2$ and shows a slight shift towards lower energies compared to stoichiometric oxides, consistent with the presence of oxygen vacancies. In the nanoscale $Hf_{1-x}Zr_xO_{2-y}$ (average size ~8 nm), such oxygen-deficient configurations are energetically favorable, particularly in the polar o-phase. The combined signatures of the reduced $t_{2g}$ intensity, the appearance of a low-energy pre-edge, and a small onset shift provide direct spectroscopic evidence for oxygen vacancy-related defect states.

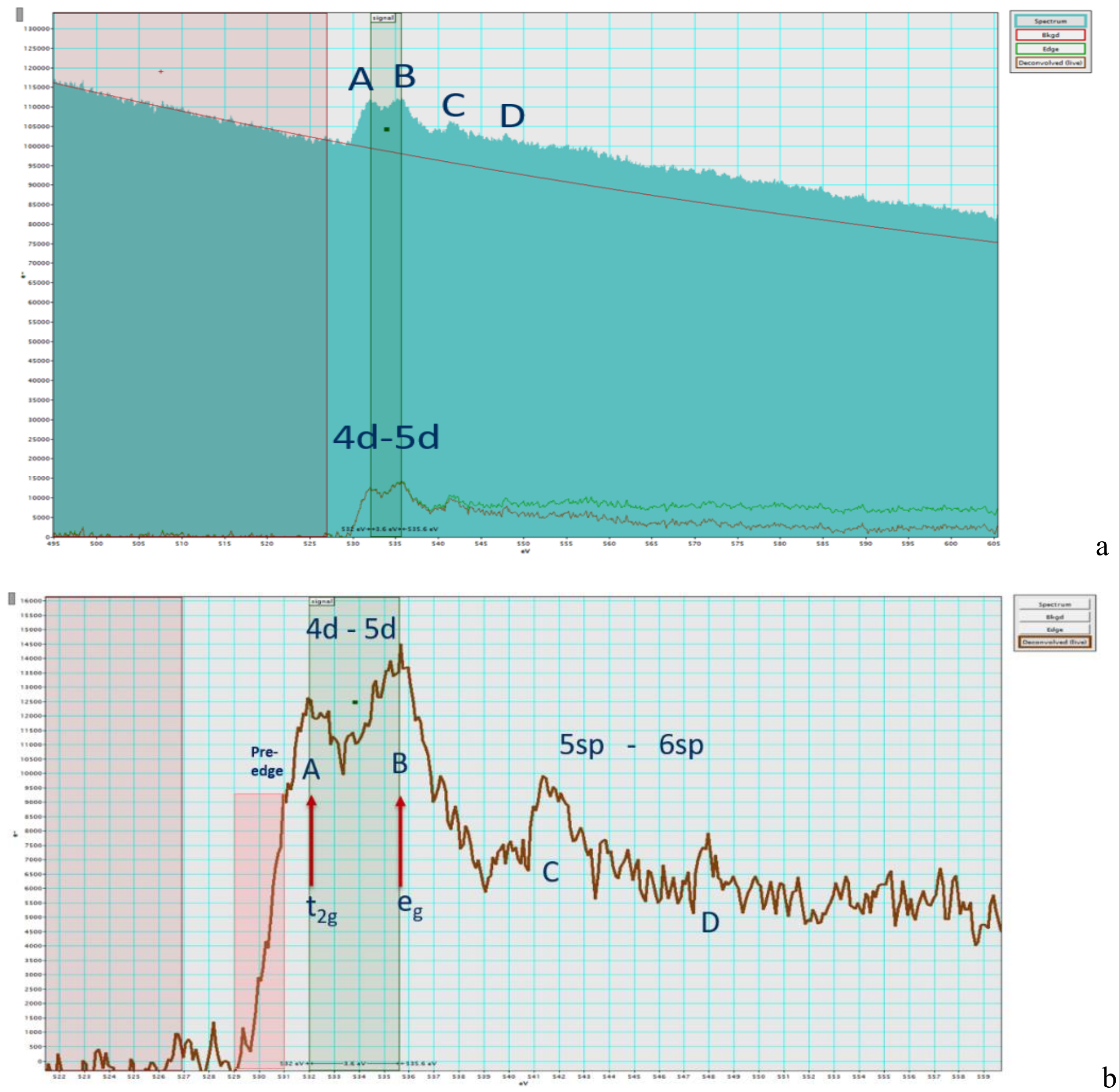

**Figure S3. (a)** The O K-edge EELS spectrum of the $Hf_{0.35}Zr_{0.65}O_{2-y}$ nanoparticles. The main features, labeled A and B, correspond to the transitions of O 1s core electrons into unoccupied O 2p states hybridized with metal $t_{2g}$ and $e_g$ orbitals, respectively. **(b)** Deconvolution of the O K-edge spectrum reveals contributions from $t_{2g}$, $e_g$, and defect-related states. The energy separation between the $t_{2g}$ and $e_g$ components defines the effective crystal-field splitting $\Delta_{oct}$. The low-energy pre-edge at ~529–531 eV, associated with oxygen vacancy-induced defect states, is shown by pink rectangle.

Although the absolute values of oxygen vacancy concentration cannot be directly quantified from the O-K edge EELS, relative changes in vacancy concentration can be inferred from the intensity of defect-related pre-edge features and the redistribution of spectral weight between the A and B peaks [79, 80, 81, 82]. The O-K edge EELS provides a sensitive probe of oxygen vacancy-related electronic structure but does not allow direct determination of absolute vacancy concentrations without external calibration (either by the first-principles calculations of DFT-based simulated spectra, or by complementary experimental techniques).

Any additional $L_{2,3}$ edges, corresponding to transition-metal impurities, such as Mn (640 – 651 eV), Fe (708 – 721 eV), or Co (779 – 794 eV) edges, are absent in the measured energy range, confirming the absence of these elements within the detection limits of the EELS (see e.g., **Fig. S4**).

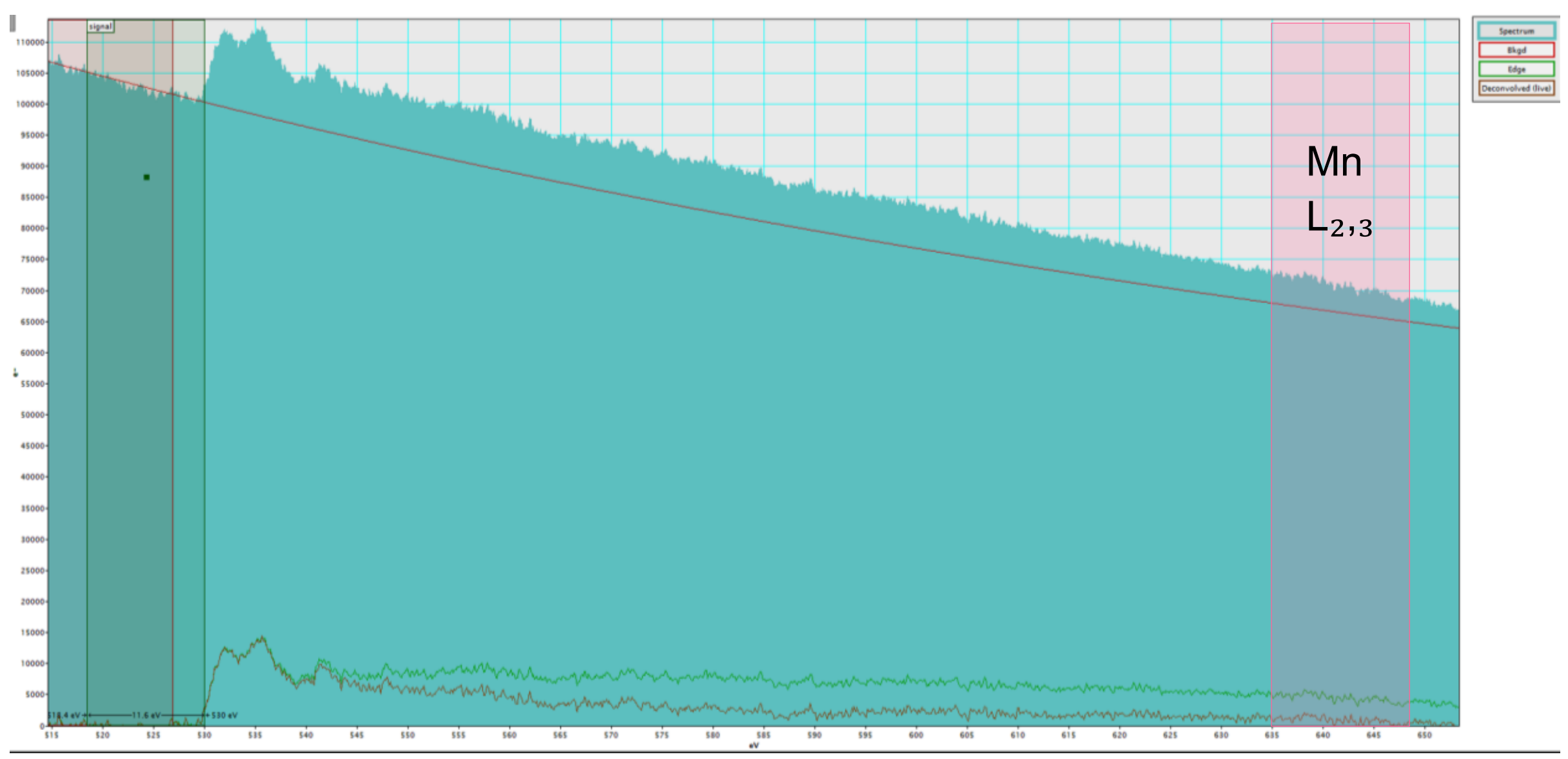

**Figure S4.** Absence of additional $L_{2,3}$ edges corresponding to transition-metal impurities such as Mn (640 - 651 eV).

The low-loss EELS region is dominated by the zero-loss peak (ZLP) followed by plasmon excitations arising from collective oscillations of valence electrons. The plasmon energy is sensitive to changes in electron density, bonding, and lattice parameters. Therefore, shifts in the plasmon peak position can serve as an additional indicator of compositional changes, defect concentration, or local structural distortions in $Hf_{1-x}Zr_xO_{2-y}$ (see **Fig. S5**).

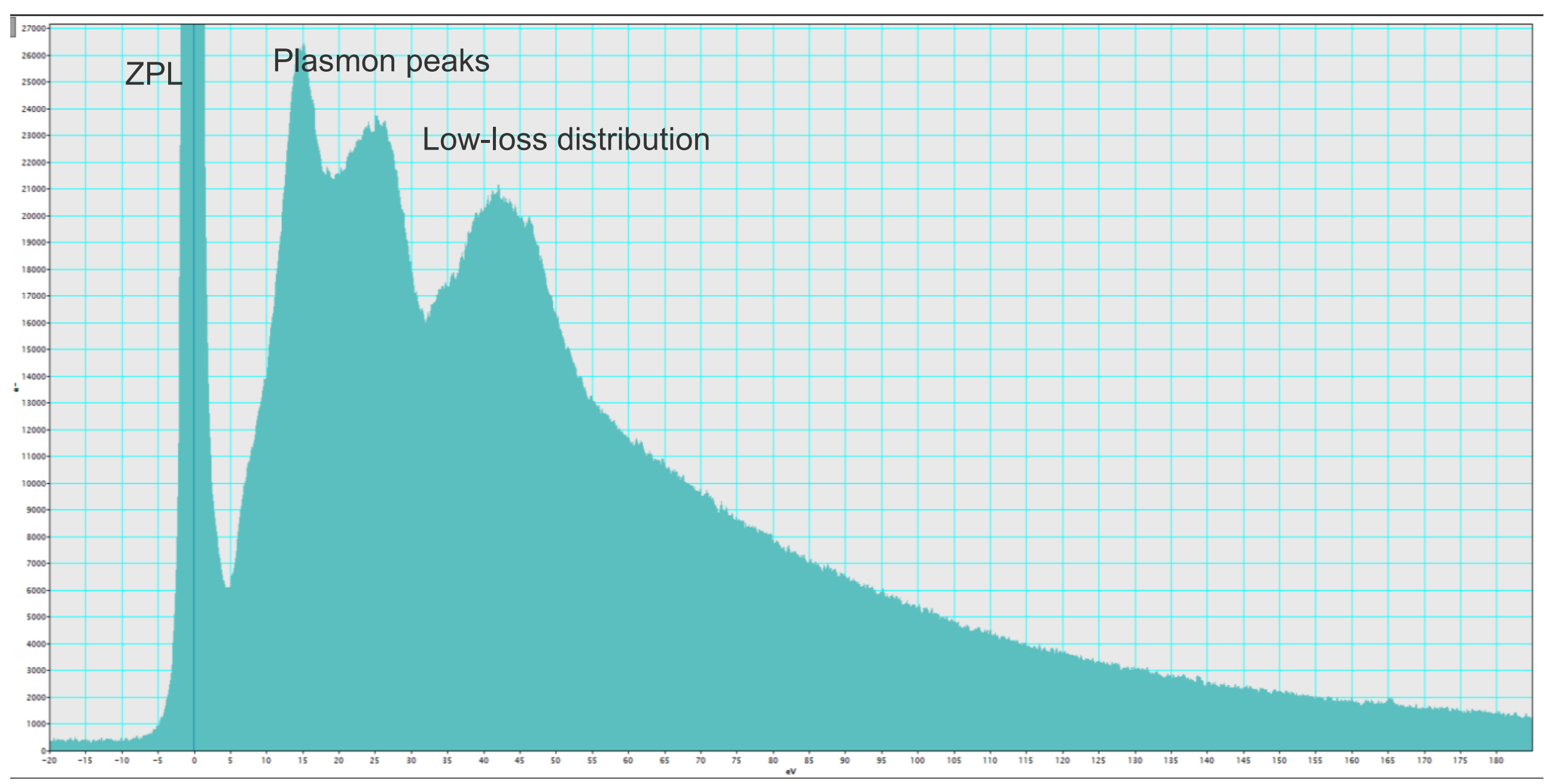

**Figure S5**. Plasmon peaks, corresponding to the valence/conduction electron density, and low-loss distribution corresponding to weakly bounded electrons.

## SUPPLEMENT S3. Raman spectra analysis

Raman spectra were recorded using a micro-Raman Renishaw ™ spectrometer equipped with a 633 nm He–Ne laser. The measurements were performed using a 50-fold magnification objective lens, with a spectral resolution of approximately 1–2 cm⁻¹. For each spectrum, three data accumulations were acquired to improve the signal-to-noise ratio. Calibration was carried out against the first-order Si peak at 520.7 cm⁻¹.

Raman spectra of $Hf_xZr_{1-x}O_{2-y}$ nanopowders, shown in **Fig. S6(a)**, exhibit a broad band in the 1000 – 3000 cm⁻¹ range, which is typical for the samples containing a high concentration of Raman-active luminescence centers, which can be associated with oxygen vacancies (see **Table S3**). The intensity of this broad feature decreases systematically with increase in zirconium content, which correlates with the stabilization of the more homogeneous o-phase and with possible reduction in the number of oxygen vacancies. For the samples with x = 0.4, the intensity of this band increases slightly in comparison with the intensity for x = 0.5, which we attributed to the increase in the m-phase fraction and the associated growth in the number of oxygen vacancy-related luminescence centers.

**Table S3.** Peculiarities of the Raman spectra of $Hf_xZr_{1-x}O_{2-y}$ nanopowders

| Sample (phase - wt. %)* | Peak position (cm⁻¹) | Assigned mode | FWHM** (cm⁻¹) | Maximal intensity (rel. units), Comment |
|---|---|---|---|---|
| $HfO_{2-y}$ o-phases - 87 % | ~1600 (broad) | defect-related luminescence | ~1900 | Maximal intensity is $13.5\cdot10^3$ (the highest concentration of vacancies) |

| Sample<br>(phase - wt. %)[*] | Peak position<br>(cm$^{-1}$) | Assigned mode | FWHM[**]<br>(cm$^{-1}$) | Maximal intensity (rel. units),<br>Comment |
|---|---|---|---|---|
| m-phase - 13 % | | | | |
| $Hf_{0.6}Zr_{0.4}O_{2-y}$<br>o-phases - 95 %<br>m-phase - 5 % | ~1600 (broad) | defect-related<br>luminescence | ~1750 | Intensity lowers to $8.1 \cdot 10^3$<br>(vacancies concentration reduces) |
| $Hf_{0.5}Zr_{0.5}O_{2-y}$<br>o-phases - 96 %<br>m-phase - 4 % | ~1550 (broad) | defect-related<br>luminescence | ~1150 | Intensity lowers further to $5.0 \cdot 10^3$<br>(vacancies concentration reduces<br>further) |
| $Hf_{0.4}Zr_{0.6}O_{2-y}$<br>o-phases - 89 %<br>m-phase - 11 % | ~1600 (broad) | defect-related<br>luminescence | ~1200 | Intensity increases slightly to<br>$5.3 \cdot 10^3$ due to the increase of the<br>m-phase fraction |

[*] According to the XRD phase analysis

[**] Full width at half maximum

No distinct phonon modes were observed below 350 cm$^{-1}$, which is consistent with the strongly disordered nanosized structure of the powders. This result agrees with the XRD data, which revealed a coexistence of o-phases and m-phases in the studied nanoparticles. The combined analysis confirms that the magnetic and dielectric properties of the $Hf_xZr_{1-x}O_{2-y}$ nanoparticles are determined by the concentration and charge state of defects, presumably oxygen vacancies, which manifest both in the Raman spectra and in the magnetic response of the studied samples.

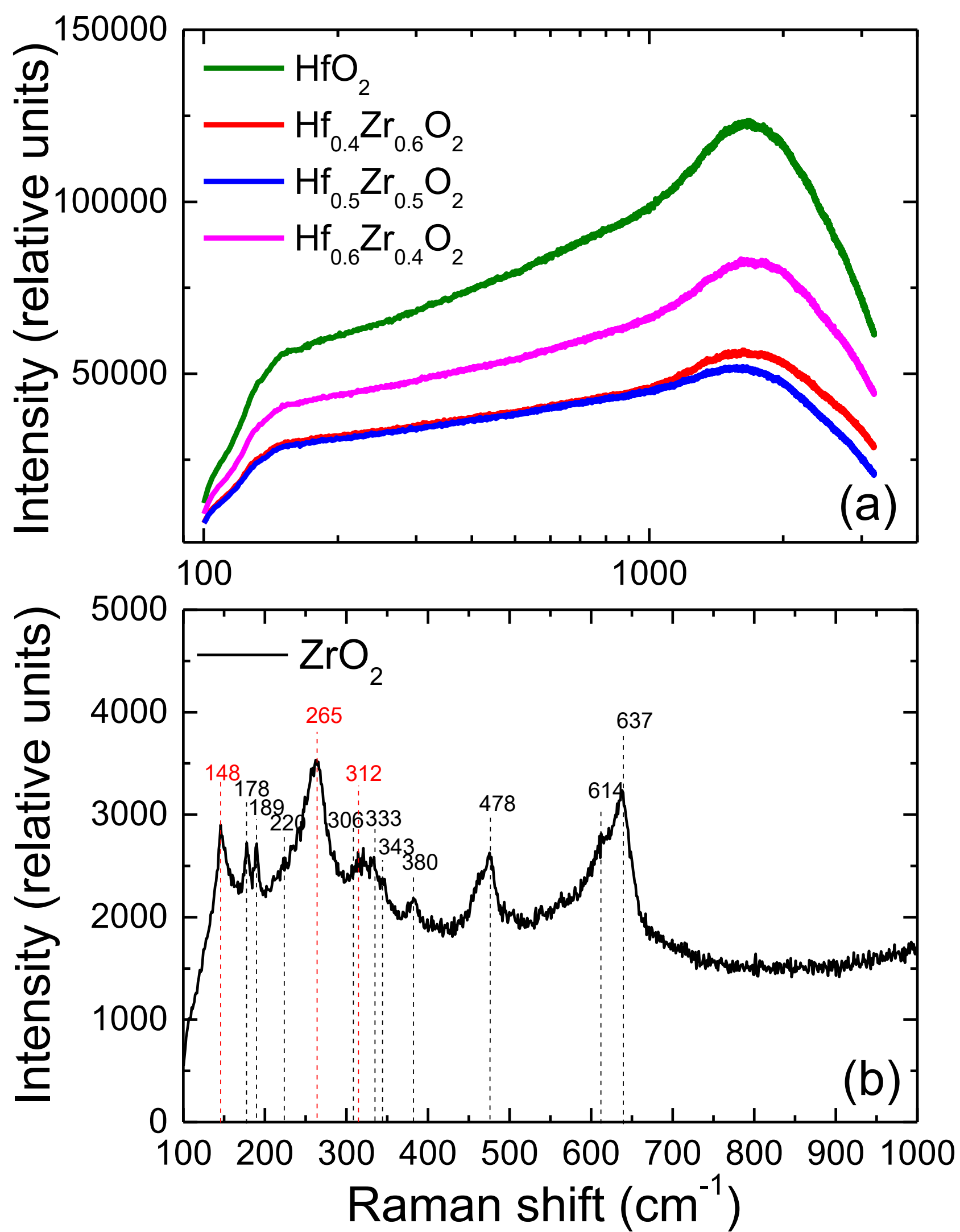

**FIGURE S6. (a)** Raman spectra of the $Hf_xZr_{1-x}O_{2-y}$ nanopowders with different hafnium content x = 1, 0.6, 0.5 and 0.4. **(b)** Raman spectrum of the commercial $ZrO_2$ nanopowder.

To identify better the crystal phases, the Raman spectra of the $Hf_xZr_{1-x}O_{2-y}$ nanoparticles were compared with the reference spectra of the commercial $ZrO_2$ nanopowders, shown in **Fig. S6(b)**. The spectrum reveals prominent bands at 148, 178, 189, 265, 478, and 637 $cm^{-1}$, with weaker features observed at 220, 306, 312, 333, 343, 380, and 614 $cm^{-1}$. The peaks at 148, 265, and 312 $cm^{-1}$ are attributed to Raman-active modes of the tetragonal phase of $ZrO_2$, corresponding to the $B_g$ mode at 148 $cm^{-1}$, $E_g$ at 265 $cm^{-1}$, and $B_g$ at 312 $cm^{-1}$. All other observed bands are associated with the m-phase of $ZrO_2$, where the $A_g$ modes appear at 178, 189, 306, 343, 478, and 637 $cm^{-1}$, and the $B_g$ modes at 220, 333, 380, and 614 $cm^{-1}$ [83]. The comparison demonstrates that, despite the broadening of Raman features due to nanosized particles, the Raman spectra of the $Hf_xZr_{1-x}O_{2-y}$ nanoparticles are consistent with a mixture of o-phases and m-phase revealed by the XRD analysis. The domination of the o-phases in the $Hf_{0.5}Zr_{0.5}O_{2-y}$ nanopowders explains the high intensity of the defect-related luminescence, whereas the increased fraction of m-phase at x < 0.5 accounts for the partial recovery of this band.

**SUPPLEMENT S4. Dependences of the fitting parameters vs. the Hf content "x"**

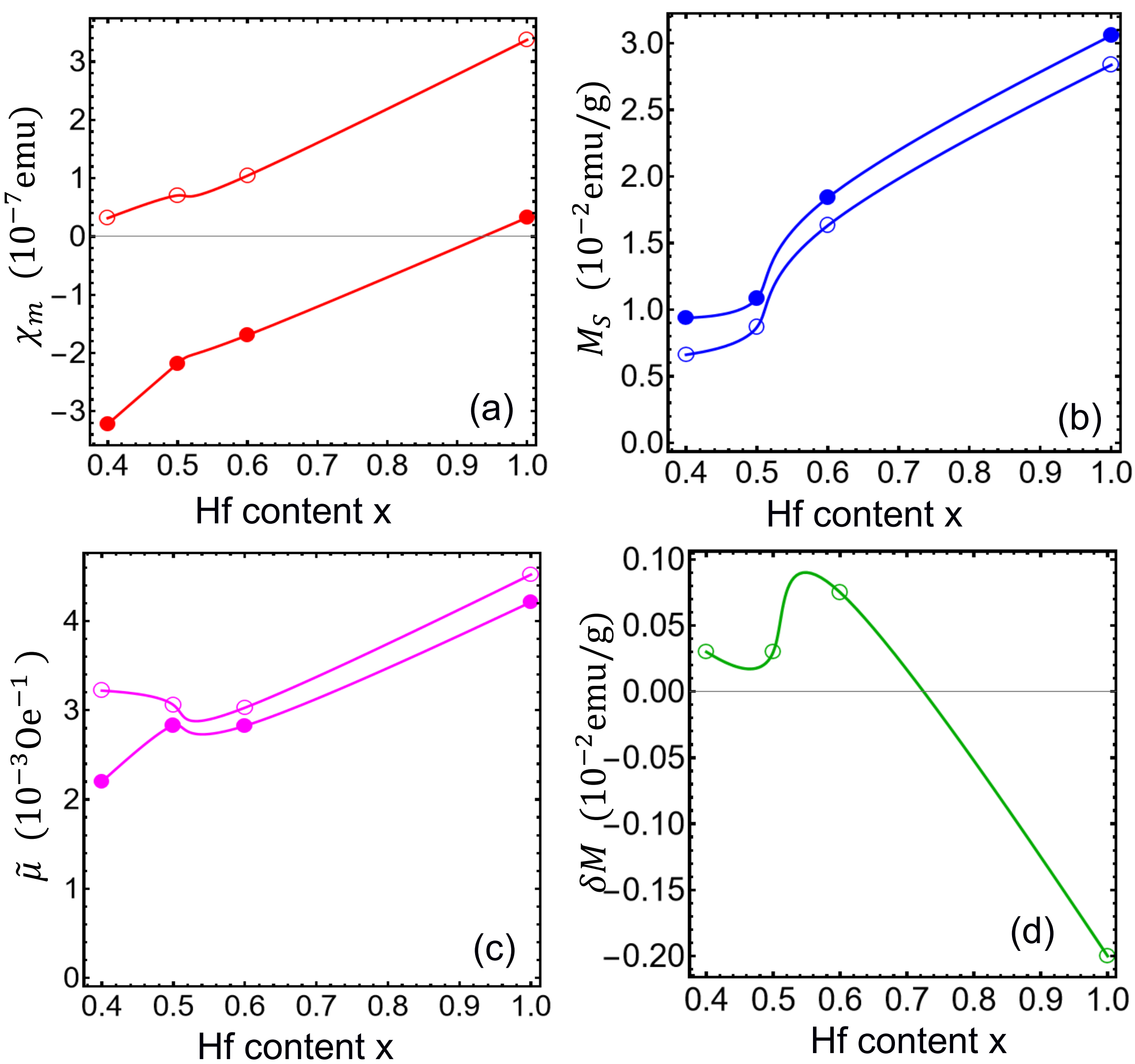

**FIGURE S7**. Dependences of the fitting parameters vs. the Hf content "x" in the $Hf_xZr_{1-x}O_{2-y}$ nanopowders. Susceptibility $\chi_m$ **(a)**, saturation magnetization $M_S$ **(b)**, magnetic dipole moment $\tilde{\mu}$ **(c)** and loop shift $\delta M$ **(d)**. Filled and empty symbols are experimentally measured values for the forward and backward direction of the magnetic field sweep, respectively.

Let us estimate the relative concentration of paramagnetic ions in the $HfO_{2-y}$ nanopowders, assuming that only $Hf^{3+}$ ions (in complex with oxygen vacancies, namely $Hf^{3+}$-$V_O$) contribute to saturated magnetization $M_S$. Magnetic moment per $Hf^{3+}$ ion is $5d^1 \rightarrow$ one unpaired electron, $S = 1/2$. The saturation magnetic moment is $\mu_{\rm sat} = gS\mu_B \approx 1\,\mu_B \cong 9.274 \times 10^{-21}$ emu (emu = erg/G in cgs units). The number of magnetic ions per gram is $N_{Hf^{3+}} = \frac{M_s}{\mu_{\rm ion}} = \frac{M_s[\frac{emu}{g}]}{9.274\times10^{-21}[emu]} = [\frac{ions}{g}]$. The number of Hf sites per gram of $HfO_2$ is $N_{\text{Hf}} = \frac{N_A}{M(\text{HfO}_2)} = \frac{6.022\times10^{23}}{210.49} \approx 2.86\times10^{21}$ Hf/g, where the molar mass $M(\text{HfO}_2) = 178.49 + 2\times16 = 210.49$ g/mol. The relative fraction (concentration) of $Hf^{3+}$ is $\frac{N_{Hf^{3+}}}{N_{\text{Hf}}} \approx 0.11$ %. Using that $M(\text{Hf}_x\text{Zr}_{1-\text{x}}\text{O}_2\,) = 32 + 178.49x + 91.22(1-x)$ g/mol, we estimated the relative concentration of paramagnetic ions in the $Hf_xZr_{1-x}O_2$ nanopowders. Results are listed in **Table S4**. Calculated fraction $\frac{N_{\text{Hf(Zr)}^{3+}}}{N_{\text{Hf(Zr)}}}$ correlates with the area of defect-related Raman spectra for all "x" except for x = 0.35 – 0.4.

**Table S4.** Relative concentration of paramagnetic ions in the $Hf_xZr_{1-x}O_2$ nanopowders

| Nominal content "x" of Hf atoms | Content "x" of Hf atoms from EDS | $M_S$ ($10^{-2}$ emu/g) (field sweep direction) | $\frac{N_{\text{Hf(Zr)}^{3+}}}{N_{\text{Hf(Zr)}}}$ (%) (calculated using nominal "x") | $\frac{N_{\text{Hf(Zr)}^{3+}}}{N_{\text{Hf(Zr)}}}$ (%) (calculated using "x" from EDS) | Area of Raman spectra (rel. units) |
|---|---|---|---|---|---|
| 0.40 | 0.35 | 0.661 (backward)<br>0.937 (forward) | 0.0187<br>0.0265 | 0.0182<br>0.0258 | $6.36\cdot10^6$ |
| 0.50 | 0.55 | 0.868 (backward)<br>1.084 (forward) | 0.0259<br>0.0324 | 0.0266<br>0.0332 | $5.75\cdot10^6$ |
| 0.60 | 0.65 | 1.631 (backward)<br>1.840 (forward) | 0.0513<br>0.0579 | 0.0526<br>0.0593 | $14.18\cdot10^6$ |
| 1.0 | 0.89 | 2.837 (backward)<br>3.058 (forward) | 0.1069<br>0.1153 | 0.1021<br>0.1100 | $25.65\cdot10^6$ |

Note that the small values of $\frac{N_{\text{Hf(Zr)}^{3+}}}{N_{\text{Hf(Zr)}}}$ ~0.1 % cannot be determined by the XPS or EELS methods. However, these values are related to the bulk of the nanoparticles and are calculated in the assumption that the dielectric/paramagnetic part of the response (proportional to $\chi_m H$) is excluded (see **Table 1** in the main text). Assuming that the factor $\Delta R$ responsible for exponential decay of defect concentration in the nanoparticle shell is about one lattice constant (~0.5 nm) and the average radius $R$ of the nanoparticles is about 10 nm, the ratio $\frac{N_{\text{Hf(Zr)}^{3+}}}{N_{\text{Hf(Zr)}}}$ is about 0.25 % (at x = 0.4) and 1 % (at x = 1) in the nanoparticle shell. The ratio increaes due to the large factor $\frac{R}{\Delta R}$.

## SUPPLEMENT S5. Resistivity measurements in the DC regime

**The first method of measurements.** Set the temperature and apply a voltage of 2 V to the sample. Wait until the current stops changing. Measure the resistance – determine the current from the voltage drop across the load resistor and divide the voltage across the sample by this current to obtain the sample resistance at each set temperature. The waiting time for the current stabilization for the first temperature point is at least 5-10 minutes. Then, without removing the voltage, we proceed to the other temperature points, etc.

**The second method of measurements.** Set the temperature and apply a voltage of 2 V to the sample and simultaneously begin measuring the current through the sample as a function of time. The measurement lasts approximately 2 minutes. In addition to integrating to determine the accumulated charge, the DC current at the end of each period of measurements is used to determine the sample resistance. This process continues at each temperature. Each measurement is performed after the sample is short-circuited through an electrical breaker for at least 2 minutes.

**Figure S8(a)** shows the resistivity measured using both methods for the $Hf_{0.5}Zr_{0.5}O_{2-y}$ sample. The black curve is the temperature dependence of the resistivity measured in the DC regime using the first method. The red curve with empty red symbols corresponds to the resistivity measured in the DC regime using the second method. Filled blue symbols are the temperature dependences of the resistivity measured in the AC regime at 4 Hz. **Figure S8(b)** is a zoomed-in part of **Fig. S8(a)**, where the differences are apparent.

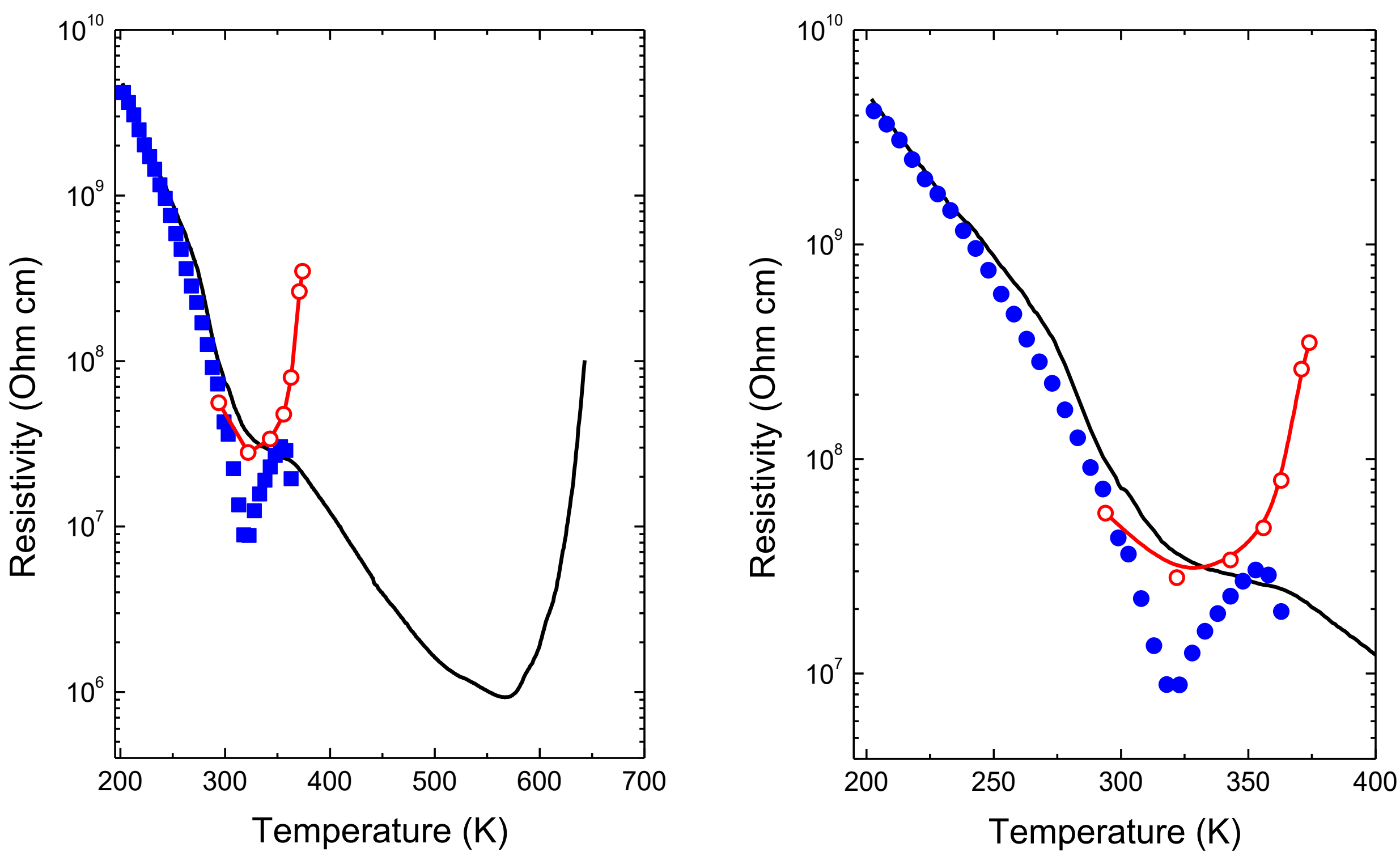

**FIGURE S8**. **(a)** The resistivity measured using both methods for the $Hf_{0.5}Zr_{0.5}O_{2-y}$ sample. The black curve is the temperature dependence of the resistance measured in the DC regime using the first method. The red curve with empty red symbols corresponds to the resistivity measured in the DC regime using the second method. Filled blue symbols are the temperature dependence of the resistivity measured in the AC regime at 4 Hz. **(b)** A zoomed-in part of **(a)**.

## References

---

[1] Roadmap on ferroelectric hafnia and zirconia-based materials and devices, APL Mater. **11**, 089201 (2023); https://doi.org/10.1063/5.0148068